\pdfoutput=1
\documentclass[10pt]{article}
\usepackage[a4paper,margin=0.95in]{geometry}
\usepackage{amsmath,amssymb}
\usepackage{booktabs}
\usepackage{graphicx}
\usepackage{hyperref}
\usepackage{xcolor}
\usepackage{enumitem}
\usepackage{caption}
\usepackage{subcaption}
\usepackage{multirow}
\usepackage[T1]{fontenc}
\usepackage{lmodern}
\hypersetup{colorlinks=true,linkcolor=blue,urlcolor=blue,citecolor=blue,
  pdftitle={An interpretable closed form for entanglement entropy from bitstrings, guided by a graph neural network},
  pdfauthor={Anas Saleh},
  pdfsubject={quant-ph},
  pdfkeywords={entanglement entropy, Rydberg atoms, graph neural network, DMRG, symbolic regression, interpretability}}
\graphicspath{{}}
\usepackage{flafter}

\title{An interpretable closed form for entanglement entropy from bitstrings,
guided by a graph neural network}

\author{Anas Saleh\\
  \small Logatta, Amman, Jordan\\
  \small \texttt{asaleh.phys@gmail.com} \quad ORCID: 0009-0006-8620-1867}
\date{}

\newcommand{\Ivar}{I}
\newcommand{\Svn}{S_{\mathrm{vN}}}

\usepackage{placeins}
\begin{document}
\maketitle

\begin{abstract}
The empirical bitstring distribution is the most accessible observable on Rydberg-atom
arrays, but the bipartite von~Neumann entropy it constrains is far costlier to obtain.
We present a six-term linear closed form for the entropy, built from bitstring features
whose choice is guided by a trained graph neural network (GNN). Probing the GNN shows that
its entropy prediction is routed through the atom pairs that straddle the bipartition
boundary, so we compute the two-point correlators over these pairs only. An exhaustive
search scored against exact entropies then selects the form. The form is human-readable
but less accurate than the GNN in distribution, where its mean absolute error (MAE) of
$0.024$~nats is $6.4$ times larger. Fit once and applied unchanged, however, it has lower
MAE than the base GNN on five of six out-of-distribution pools and ties it on the sixth.
Refit per Hamiltonian family, it reaches $29$--$70$~millinats (mnat) on the four
sign-definite families we tested. Further out of distribution, we test it on
density-matrix renormalisation group ground states of up to one hundred atoms. There the
slopes must be refit per size, which gives a cross-validated MAE of $15$--$37$~mnat, and
size laws fitted to the slopes without the target size give $24$--$95$~mnat with no labels
at that size.
\end{abstract}

\section{Introduction}\label{sec:intro}

The bipartite von Neumann entropy $\Svn(\rho_A)\!=\!-\mathrm{Tr}\,\rho_A\log\rho_A$ of a
subsystem $A$ separates area-law from volume-law phases~\cite{EisertCramerPlenio2010},
detects topological order~\cite{JiangWangBalents2012}, and measures how far an analogue
quantum simulator has run beyond the reach of classical methods~\cite{Shaw2024}. Among
such simulators, programmable Rydberg-atom arrays realise $Z_k$
crystalline phases, quantum critical points, and tunable disorder at intermediate system
sizes~\cite{Bernien2017,Ebadi2021,Scholl2021}. Their most accessible observable is the
empirical bitstring distribution. Each measurement in the occupation basis returns one
length-$N$ bitstring, so repeated shots accumulate an empirical distribution over the
$2^N$ configurations. The entanglement entropy, however, follows from the quantum state itself and
is far costlier to obtain (\S\ref{sec:background}).

\begin{figure}[!ht]
\centering
\includegraphics[width=0.98\textwidth]{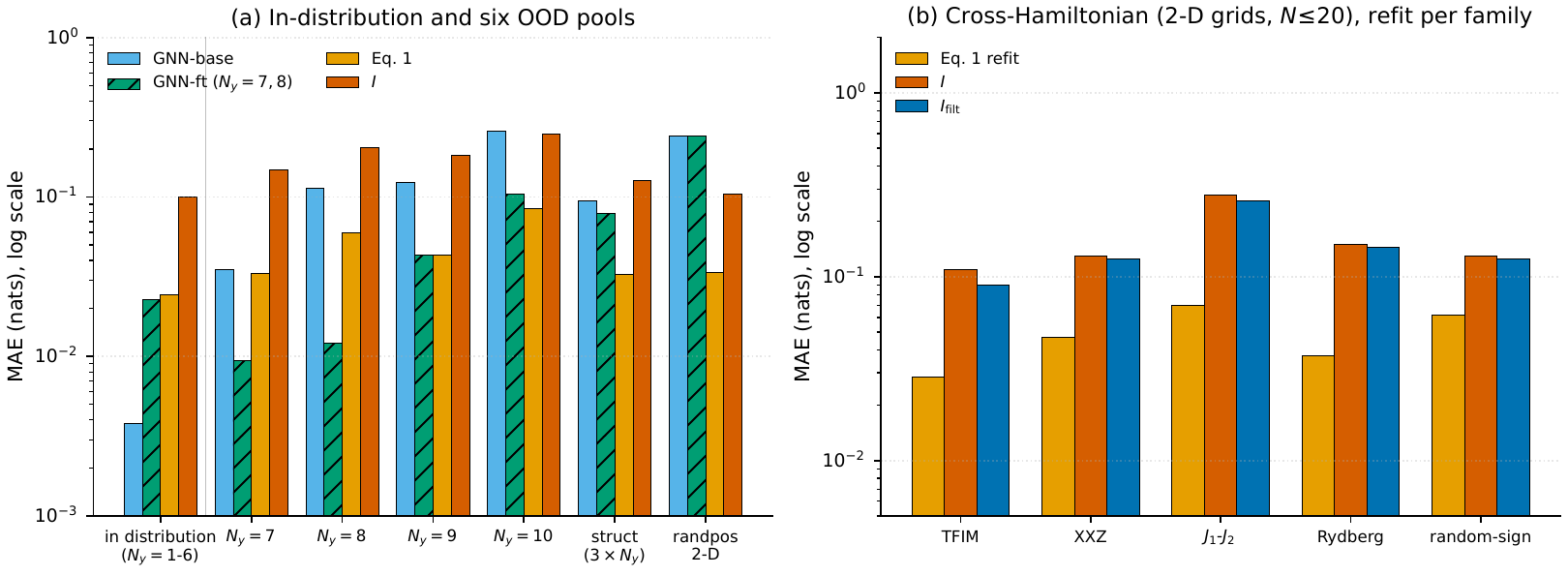}
\caption{Headline summary. \emph{(a)} MAE on the in-distribution evaluation set and the
six OOD pools for the base GNN (\cite{Saleh2025}, sky blue), the
fine-tuned GNN checkpoint (bluish-green, \emph{hatched}), the six-feature closed form
Eq.~\ref{eq:1} (orange), and the classical mutual information $\Ivar$ (vermilion), on a
log scale. The fine-tuned column is the published checkpoint fine-tuned
in~\cite{Saleh2025} for the $N_y\!=\!7,8$ ladders and applied unchanged. Pools right of
the separator are OOD relative to the GNN's training distribution. Eq.~\ref{eq:1} has
lower MAE than the base GNN on five OOD pools and ties it on the sixth ($N_y\!=\!7$).
The fine-tuned GNN checkpoint has the lowest MAE on the size-OOD pools at
$N_y\!\in\!\{7,8\}$ and ties Eq.~\ref{eq:1} at $N_y\!=\!9$, but has higher MAE than
Eq.~\ref{eq:1} at $N_y\!=\!10$ and on the geometry-shifted pools. \emph{(b)} Refit-per-family MAE on five
Hamiltonian families (2-D grids, $N$ up to $20$): the same six-feature form has lower
MAE than both $\Ivar$ and $\Ivar_{\mathrm{filt}}$~\cite{Asad2024} on every family
($\Ivar$ values in Tables~\ref{tab:ood_mae} and~\ref{tab:cross_ham_indist};
$\Ivar_{\mathrm{filt}}$ in SI Table~\ref{tab:crossham_ext}). The per-family GNN is omitted because it would need
re-featurising and retraining for each family. The same architecture reaches $2.6$~mnat on transverse-field Ising configurations in~\cite{SalehThesis}
(\S\ref{sec:gen_indist}).}
\label{fig:headline}
\end{figure}

The occupation-basis distribution constrains the entropy without determining it, so estimating $\Svn$ from bitstrings is an inverse problem that can only be solved approximately. Existing estimators sit at two extremes of an
interpretability--accuracy trade-off (\S\ref{sec:background}). The analytic baselines
built from the classical mutual information $\Ivar$ are closed-form and human-readable,
but $\Ivar$ underestimates $\Svn$, with a mean absolute error (MAE) of $\approx\!0.10$~nats on the evaluation set of \S\ref{sec:setup}. At the other extreme, the graph neural network (GNN) of~\cite{Saleh2025} ($19.9$M
parameters) reaches $\sim\!4$~millinats (mnat), but it is a black box whose MAE grows out
of distribution. These numbers are for its base checkpoint; \S\ref{sec:ood} also uses one
fine-tuned for longer ladders. Between these extremes, we aim for an estimator that (i)~can be fit once
and applied unchanged across size and geometry shifts and (ii)~is robust against shot noise.
Its form should also (iii)~be simple enough to check by hand and to carry to a new
system without looking inside a trained network.

To build it, we let the GNN guide the featurisation and then select and test the closed form
against exact entropies without the network. Probing the trained GNN~\cite{Saleh2025}
shows that its $\Svn$ prediction is routed through the boundary set
$E_b\!=\!A\!\times\!B$, the atom pairs that straddle the cut (\S\ref{sec:gnn_boundary}).
We therefore compute the two-point correlator features over the boundary pairs and leave out the
bulk pairs ($A\!\times\!A$, $B\!\times\!B$) that a default featurisation would include. An
exhaustive search over a shortlist that includes these features, scored against exact entropies,
selects a six-term linear closed form, Eq.~\ref{eq:1} (\S\ref{sec:closed}). To
test the restriction, we rerun the search with the bulk correlators added to the
shortlist. Its optimum is still boundary-only (\S\ref{sec:panel}), so per feature the
bulk correlators carry less signal than the boundary ones. The analytic baselines in the
literature, $\Ivar$ and its filtered variant~\cite{Asad2024}, contain no such
boundary-indexed correlator terms, because they are global functionals of the bitstring
distribution.

\paragraph{The three contributions.} \textbf{(1)~A GNN-guided closed form}
(\S\ref{sec:gnn_boundary}--\ref{sec:eval}). Eq.~\ref{eq:1} is a human-readable form with
six features and no hyperparameters. Fit once on $2\!\times\!N_y$ Rydberg ladders and
applied unchanged, it has lower MAE than the base GNN on five of the six out-of-distribution (OOD) pools and ties
it on the sixth. Under shot noise it has lower MAE than the classical mutual information
above a crossover shot count (\S\ref{sec:shot}).

\textbf{(2)~Transfer across Hamiltonians and sizes} (\S\ref{sec:generality},
\S\ref{sec:dmrg}). Refit per family, the same six features give lower MAE than the
analytic baselines on every family we tested. We also refit the form per size against
density-matrix renormalisation group (DMRG) data on the balanced half cut of the ladders.
These fits extend the closed form to $N\!=\!100$ with a
cross-validated MAE of $15$--$37$~mnat. Across these refits, two of the six slopes follow $1/N$ laws, one grows and curves downward,
and the other three show no resolvable trend.

\textbf{(3)~A mechanistic case study of the GNN} (\S\ref{sec:decod}). Causal
interventions on the GNN show that its prediction is routed through the boundary pairs
(\S\ref{sec:gnn_boundary}), which is the finding behind~(1). Decoding analyses show that the GNN encodes the quantities of Eq.~\ref{eq:1} redundantly, over many correlated directions, so a closed form read off the network is opaque or less accurate. Eq.~\ref{eq:1}, whose fit uses none of the network's weights, outputs or activations, is not a distillation of the network.

Across Hamiltonian families and sizes, the six features carry over and only the slopes are refit. Two opaque models are more accurate than Eq.~\ref{eq:1}. A gradient-boosted tree model (XGBoost~\cite{ChenGuestrin2016}) on the $25$-feature shortlist, trained on the full training set, has lower MAE in distribution and on all six OOD pools. The GNN, which reads the full correlation graph, has lower MAE still in distribution. Eq.~\ref{eq:1} gives up this accuracy for a form that can be read and checked by hand.
\section{Background and related work}\label{sec:background}

\subsection{Rydberg-atom arrays}
A Rydberg array traps individual neutral atoms in optical tweezers and laser-couples each
to a high-lying Rydberg state at Rabi frequency $\Omega$ and detuning $\Delta$. Atoms
in the Rydberg state repel through a van der Waals interaction, which gives the
Hamiltonian of \S\ref{sec:setup}~\cite{BrowaeysLahaye2020}. Within the blockade radius
$R_b$, the distance at which this interaction equals the Rabi frequency, double excitation is suppressed.
Sweeping the dimensionless controls $(\Delta/\Omega,\,R_b/a)$, with $a$ the lattice
spacing, carries the array through the $Z_k$ crystalline phases
and the quantum critical points between them. Along such sweeps, the ground states range from near-product
to highly entangled~\cite{Semeghini2021}. The most entangled already lie beyond exact
classical simulation~\cite{Shaw2024}, so the arrays are also a test case for classical
estimators of their properties.

\subsection{Entanglement entropy and its estimation}
For a pure global state $|\psi\rangle$ the bipartite von Neumann entropy is
$\Svn=-\mathrm{Tr}\,\rho_A\log\rho_A$, with $\rho_A=\mathrm{Tr}_B\,|\psi\rangle\!\langle\psi|$.
It is a nonlinear functional of the state and is fixed by no single measurement basis,
so it is costly to measure. Full tomography scales exponentially in $N$, while tomography-free
protocols trade that cost for overhead in copies, gates, or post-processing. These protocols include
two-copy interference, proposed in~\cite{DaleyPichler2012,AbaninDemler2012} and realised
in~\cite{Islam2015,Kaufman2016}, randomised
measurements~\cite{ElbenPRA2019,Brydges2019,ElbenNRP2023}, and classical
shadows~\cite{Huang2020}. The cheapest signal, by contrast, is the occupation-basis bitstring
distribution that any projective readout already returns. This distribution does not
determine $\Svn$ (two states with the same diagonal distribution can differ in
entanglement), but it does bound it. The classical mutual information
$\Ivar\!=\!H(X_A)\!+\!H(X_B)\!-\!H(X_A,X_B)$ of the cut bitstrings lower-bounds $\Svn$
through the Holevo bound (for a pure state, $\Svn\!\ge\!\Ivar$~\cite{HolevoBound,WildeBook}),
and filtering low-probability bitstrings tightens the estimate~\cite{Asad2024}. These
analytic baselines are closed-form and readable, but they underestimate $\Svn$ on our pools (Tables~\ref{tab:in_dist_mae} and~\ref{tab:ood_mae} for $\Ivar$, SI Table~\ref{tab:crossham_ext} for $\Ivar_{\mathrm{filt}}$).

\subsection{Machine-learned predictors and interpretability}
Neural quantum states represent many-body wavefunctions variationally~\cite{Carleo2017_NQS},
and neural networks classify phases directly from configurations~\cite{Carrasquilla2017};
Ref.~\cite{CarleoRMP2019} reviews the methodology. Neural networks have also been applied
to entanglement. They map measured moments to an entanglement measure with far fewer
measurements than tomography~\cite{Gray2018}, reconstruct many-body states from
measurement data~\cite{Torlai2018}, and quantify entanglement from incomplete measurement
settings~\cite{Koutny2023}. These predictors are accurate but opaque, and none gives a
closed form. For the entanglement entropy, the GNN of~\cite{Saleh2025,SalehThesis} reads
the correlation graph built from the bitstrings and predicts $\Svn$; it is the network we
probe in this work.

Two lines of work aim instead at transparent descriptions. Symbolic regression searches for
closed-form expressions. Some methods fit data directly, beginning with evolutionary
search over experimental data~\cite{SchmidtLipson2009}, followed by sparse regression over
candidate terms~\cite{Brunton2016} and AI~Feynman, which recovers physical laws from
data~\cite{UdrescuTegmark2020}. Others distil a trained network. One such study extracts
symbolic force laws from the learned messages of a graph network~\cite{Cranmer2020}.
Another rediscovers Newtonian gravity from a graph network trained on solar-system
trajectories~\cite{Lemos2023}. Mechanistic interpretability instead probes a fixed network
with causal interventions to identify the intermediate variables its computation relies
on~\cite{Geiger2021}. We combine probing with exhaustive search. Probes of the trained GNN
decide which atom pairs the correlator features are computed over
(\S\ref{sec:gnn_boundary}), and we then select the closed form by exhaustive search
(\S\ref{sec:closed}).
\section{Setup}\label{sec:setup}

\paragraph{Hamiltonian.} In units of $\Omega=1$,
\[ H = \tfrac12\sum_i\sigma^x_i - (\Delta/\Omega)\sum_i n_i
     + \!\!\sum_{i<j,\,r_{ij}\le R_{\mathrm{cut}}}\!\! (R_b/a)^6/r_{ij}^6\,n_i n_j,
   \qquad n_i=(1-\sigma^z_i)/2. \]

\paragraph{Conventions.} A $2\!\times\!N_y$ ladder has $N\!=\!2N_y$ atoms, and distances
$r_{ij}$ are in units of the lattice spacing $a$. Entropies and mutual informations are
in nats, and errors in nats or mnat ($10^{-3}$~nats). All target entropies $\Svn$ in the pools below are
exact-diagonalisation (ED) ground-state values; the $N\!\ge\!24$ scaling study of
\S\ref{sec:dmrg} uses DMRG instead. Every feature is computed from the exact distribution $|\psi|^2$ (finite shots in \S\ref{sec:shot}). The classical mutual information of the cut $A|B$
is $\Ivar$, and
$\Ivar_{\mathrm{filt}}$ is the filtered mutual information of Kaufman
et al.~\cite{Asad2024}.

\paragraph{Pools.} We use a $6{,}000$-row training set ($1{,}000$ per
$N_y\!\in\!\{1,\dots,6\}$) and a $30{,}000$-row evaluation set ($5{,}000$ per
$N_y$), both reservoir-sampled, with parameters drawn uniformly over
$\Delta/\Omega\!\in\![0,6]$ and $R_b/a\!\in\![0.1,4]$. Six OOD pools of $1{,}500$
rows each lie beyond the training distribution: $N_y\!\in\!\{7,8,9,10\}$ ladders
(the size-OOD pools), a $3\!\times\!N_y$ structural pool (``struct'') and a
random-2-D-position pool (``randpos'', $N\!\in\!\{6,\dots,18\}$). Struct and randpos are the geometry-shifted pools. The OOD pools extend $R_b/a$ to $5$, and the interaction cutoff is $R_{\mathrm{cut}}\!=\!6a$ throughout.
Bipartition masks are uniformly random in every pool, so contiguous and
non-contiguous cuts both occur. Every row of the two sets carries a unique
$(N_y,R_b/a,\Delta/\Omega)$ triple, so the evaluation uses Hamiltonians unseen in
training; it does not test interpolation across masks on shared ground
states.

\paragraph{Architecture and checkpoints.} The GNN~\cite{Saleh2025} has
$19{,}866{,}119$ parameters and hidden dimension $512$. After a node encoder, its six message-passing
layers, each with a learned per-edge gate, alternate \texttt{GINEConv} \cite{Hu2020GINE} (even depths $0,2,4$, each followed by a per-node
refinement block) and \texttt{TransformerConv}~\cite{Shi2021TransformerConv}
($8$ heads, odd depths $1,3,5$). We write GINEConv-$k$ and TConv-$k$ for the layer at
depth $k$. Node features are $(x,y,\langle n_i\rangle,\text{mask})$, and edge features
are $(\theta_{ij},c_{ij},r_{ij}/\sqrt N)$ (orientation, two-point correlator
$c_{ij}=\langle n_in_j\rangle-\langle n_i\rangle\langle n_j\rangle$, scaled distance) on
the complete graph, with one undirected edge per pair ($i<j$). The readout is a
dual-head \texttt{Set2Set}~\cite{Vinyals2016Set2Set}. Both checkpoints we use are published artefacts
of~\cite{Saleh2025}: the base GNN (``base'' or GNN-base) and the checkpoint
fine-tuned there for $N_y\!=\!7,8$ (``fine-tuned'' or GNN-ft).
We retrain neither.
\section{What the GNN reads: boundary routing}\label{sec:gnn_boundary}

We ask which atom pairs the trained GNN of~\cite{Saleh2025} relies on to predict $\Svn$.
The boundary set $E_b=A\!\times\!B$ holds the pairs that straddle the bipartition, however
far apart, and the bulk holds the pairs with both atoms on the same side of the bipartition. The probes show that the prediction is routed through the boundary set. This finding sets the feature restriction of \S\ref{sec:closed}.

\paragraph{Design.} We run three probes. Probes~(i) and~(ii) isolate the bipartition with a
two-mask comparison, which evaluates the GNN twice on the same Hamiltonian under two
independent uniformly random bipartition masks $m,m'$. Atom positions, correlators
and graph are shared, so any difference comes from the bipartition alone. Probe~(iii)
instead compares boundary-edge deletion with bulk and count-matched random deletion. We
report quantities per layer (names as in \S\ref{sec:setup}). SI~\S\ref{app:probes} gives the
per-layer tables for all three probes, with the bulk- and random-edge controls at every layer; the text quotes
these controls for the GINEConv layers only.

\paragraph{(i) Attention routing.} At each TransformerConv layer we head-average the
per-edge attention weight $\bar\alpha_{ij}$ and track its change on edges whose
boundary status flips between $m$ and $m'$ (they cross the cut under one mask
only). The boundary-flip shift $\Delta\bar\alpha$ is the mean over these edges of
$\bar\alpha$ when the edge is boundary minus $\bar\alpha$ when it is bulk. It is
$+0.204$ at TConv-3, about $70\times$ its value at TConv-1 and $12\times$ at TConv-5.
An independent ordinary least squares (OLS) regression of $\bar\alpha$ on eight
per-edge features gives the boundary indicator a partial $R^2$ of $13.8\%$ at
TConv-3, $3\times$ that of the next feature (normalised inter-atom distance). TConv-3
thus re-routes attention by the bipartition, but attention weights are debated as
explanations~\cite{JainWallace2019,WiegreffePinter2019}. The routing conclusion
therefore rests on the causal interventions of (ii)--(iii), and the attention
statistics corroborate it.

\paragraph{(ii) Activation patching.} To find which layers carry the bipartition
forward, we use activation patching~\cite{Meng2022ROME}: we overwrite a layer's output on the $m'$ pass with its $m$-pass value and
run the remaining layers. The patching ratio $R=(\hat S^{\mathrm{patched}}-\hat
S_{m'})/(\hat S_m-\hat S_{m'})$ (hats mark GNN predictions, subscripts the mask) is
the fraction of the bipartition-induced prediction shift recovered by that single
substitution. We find $R\!\ge\!0.9$ for TConv-1 and TConv-3, but $\le\!0.18$ for
every GINEConv layer.

\paragraph{(iii) Edge removal.} We delete a class of edges (boundary, bulk, or a
count-matched random set) from a single layer's message passing and report MAE
relative to the unmodified baseline. Every layer weights its edge inputs by a learned
per-edge gate, and we rescale the surviving gate weights to keep their total fixed. At
TransformerConv layers the softmax attention then runs over the surviving edges only. At the GINEConv layers, deleting the boundary edges inflates MAE by
$16.6/10.7/21.8\times$ (in depth order), $2.3$--$3.4\times$ the count-matched
random-edge control ($7.1/3.1/7.9\times$). Bulk deletion also inflates MAE
($4.7$--$7.6\times$), so graph damage alone is costly, but boundary deletion costs
much more. TConv-1 is essentially insensitive to which edges are present, consistent
with node features that already carry the bipartition.

\paragraph{Synthesis.} Together, the probes suggest a division of labour across layers.
GINEConv-0 already relies on the boundary edges, since deleting them costs more than
deleting bulk or random ones. TConv-1 forwards the bipartition with
distance-weighted, bipartition-agnostic attention, TConv-3 re-routes by the boundary,
and later GINEConv layers and TConv-5 re-read the boundary edges
(SI Table~\ref{tab:probe_patch_remove}). Because the input node features
already include the bipartition mask, no probe isolates how the bipartition enters the
node states. The rest of
the paper uses only the empirical result that the prediction depends on which edges sit
on the $A\!\mid\!B$ boundary. The probes point to the boundary pairs $E_b$, so the
correlator features of the candidate panel in \S\ref{sec:panel} are computed over $E_b$.
A bulk-inclusive search in \S\ref{sec:panel} tests this restriction independently.
\section{The closed form}\label{sec:closed}

With the correlator features restricted to the boundary, an exhaustive search over
six-feature subsets of a curated feature shortlist (\S\ref{sec:exhaustive}) selects a linear closed form, Eq.~\ref{eq:1}. Each subset is fit to exact labels and uses none of the network's weights, outputs or activations, so the GNN enters Eq.~\ref{eq:1} only through the boundary restriction.

\subsection{The candidate panel and the boundary restriction}\label{sec:panel}

We start from a panel of $93$ candidate statistics of the bitstring distribution and the bipartition mask, sorted into eight groups. The information-theoretic baselines
($\Ivar$, Shannon/conditional entropies, R\'enyi entropies of orders $2$, $3$ and
$\infty$, participation ratio) form a group of their own. The other groups hold
marginal-density statistics, effective-dimension/geometry counts, boundary-correlator
kernel sums, extremal-correlator statistics, sign-distribution counts, cross-product/ratio
features and R\'enyi-2 analogues. We search the panel in
two ways. (i)~A greedy forward OLS over the full panel adds, at each step, the
feature that most lowers training-set MAE. Its six-feature result, the greedy form,
has in-distribution MAE $25.4$~mnat and OOD-mean MAE $52.8$~mnat (the mean over the
OOD pools of \S\ref{sec:setup}). (ii)~Scoring all
$\binom{93}{6}\!\approx\!7 \times 10^8$ subsets of the full panel is infeasible, so we
curate a $25$-feature shortlist (App.~\ref{app:shortlist}) and search all
$\binom{25}{6}$ of its subsets (\S\ref{sec:exhaustive}). The best,
Eq.~\ref{eq:1}, reaches $24.4/47.8$~mnat on the same two measures. Of the two searches,
the exhaustive search gives the lower MAE on both measures, so curating the shortlist
costs no accuracy relative to the full-panel greedy search.

The boundary restriction comes from
the boundary routing of the GNN (\S\ref{sec:gnn_boundary}), so we test it independently
with a bulk-inclusive search. This search repeats the exhaustive search with the bulk
correlators added to the boundary ones, on $6{,}000$ training rows and $8{,}000$ held-out
rows drawn at random from the evaluation set. The optimum is still boundary-only and reproduces Eq.~\ref{eq:1}. The first subset containing a bulk correlator ranks
$37$th, and forcing one into the six-feature set raises MAE by ${\approx}1$~mnat
(paired-bootstrap CI $[0.58,1.35]$~mnat). Adding bulk correlators to Eq.~\ref{eq:1},
however, lowers held-out MAE slightly (SI~\S\ref{app:bulk}), so the bulk
carries some signal,
but the boundary carries more per feature. This per-feature advantage is consistent
with the boundary routing found by the causal probes of \S\ref{sec:gnn_boundary}.

To measure what dropping the restriction costs, we keep the same six features but
compute their correlator terms over all atom pairs. MAE then rises by $4.2$~mnat ($17\%$) in distribution and by
$8.9$~mnat ($27\%$) on the struct geometry (bootstrap CIs separated;
SI~\S\ref{app:bulk}).

\subsection{Exhaustive search over the shortlist}\label{sec:exhaustive}

The exhaustive search covers all $\binom{25}{6}\!=\!177{,}100$ six-feature subsets of
the $25$-feature shortlist (App.~\ref{app:shortlist}). We fit each subset by
no-intercept OLS on the training set and rank them by MAE on the evaluation set,
which is held out from the fit. The top-ranked six-feature form is
\begin{equation}\label{eq:1}
\boxed{\hat \Svn =
  b_1 \log(\Ivar\!+\!1)
+ b_2\!\sum_{E_b}\!\sqrt{|c_{ij}|}
+ b_3\!\sum_{E_b}\!\tanh|c_{ij}|
+ b_4(\max_{E_b}|c_{ij}|)^2
+ b_5\min_{E_b} c_{ij}
+ b_6\min(H_A,H_B)\,\Ivar}
\end{equation}
where $c_{ij}$ are the two-point correlators of \S\ref{sec:setup}, with sums and extrema
over $(i,j)\!\in\!E_b$. The summed single-site entropy of region $X$ is
$H_X=\sum_{i\in X}h_2(\langle n_i\rangle)$, with $h_2(x)=-x\log x-(1\!-\!x)\log(1\!-\!x)$.
This sum of marginal entropies differs from the joint entropy $H(X_A)$ of \S\ref{sec:background}.
Eq.~\ref{eq:1} also ranks first by the score (in-distribution plus OOD-mean MAE). Its $\arctan$ twin, with the $\tanh|c|$ kernel replaced by $\arctan|c|$, ranks
second, tied in in-distribution MAE to four decimals (SI Table~\ref{tab:top5}), and the top-$50$ forms lie within $0.003$~nats of one another in MAE. Five of the six features of the greedy form
(\S\ref{sec:panel}) are also in Eq.~\ref{eq:1}, and the sixth is the boundary sum of
$c^2$, in place of the $\tanh|c|$ kernel sum. The fitted slopes are stable, with a maximum
per-slope coefficient of variation (CV) of $5.2\%$ over $B\!=\!2{,}000$ bootstrap
resamples (Table~\ref{tab:eq1_slopes}). Eq.~\ref{eq:1} is $A\!\leftrightarrow\!B$ symmetric by construction, and its non-negativity is verified numerically (SI~\S\ref{app:struct}).

Searching beyond linear forms, we ran symbolic regression (PySR~\cite{PySR2023}) on the
shortlist features, with
an MAE objective and $150$ iterations. On the held-out rows of SI~\S\ref{app:pysr}, its best expression reaches
$0.021$~nats,
against $0.024$~nats for Eq.~\ref{eq:1} and $0.022$~nats for the $25$-term linear
fit. It gets there only as a complexity-$30$ nested expression that is hard to read. XGBoost on
the same features reaches $0.010$~nats on these rows. Of these models, all those with lower MAE
than Eq.~\ref{eq:1} are harder to read (PySR configuration in SI~\S\ref{app:pysr}).

\begin{table}[!ht]\centering\small\renewcommand{\arraystretch}{1.1}
\begin{tabular}{lccc}
\toprule
slope & training-set fit & BCa $95\%$ interval & boot.\ CV~\% \\
\midrule
$b_1$ ($\log(\Ivar\!+\!1)$)        & $+1.357$ & $[+1.327,+1.391]$ & $1.2$ \\
$b_2$ ($\sum_{E_b}\sqrt{|c|}$)     & $+0.080$ & $[+0.076,+0.085]$ & $2.9$ \\
$b_3$ ($\sum_{E_b}\tanh|c|$)       & $-0.240$ & $[-0.253,-0.228]$ & $2.7$ \\
$b_4$ ($(\max_{E_b}|c|)^2$)        & $-6.379$ & $[-6.725,-6.030]$ & $2.8$ \\
$b_5$ ($\min_{E_b} c$)             & $-1.522$ & $[-1.601,-1.442]$ & $2.7$ \\
$b_6$ ($\min(H_A,H_B)\,\Ivar$)     & $+0.114$ & $[+0.103,+0.126]$ & $5.2$ \\
\bottomrule
\end{tabular}
\caption{Eq.~\ref{eq:1} slopes (no-intercept OLS on the training set). BCa~\cite{Efron1987BCa}
(bias-corrected and accelerated) $95\%$ intervals and per-slope CV, from the same
$B\!=\!2{,}000$ bootstrap resamples.}\label{tab:eq1_slopes}
\end{table}

\subsection{Reading the six features}\label{sec:why}

\paragraph{Information-theoretic pair.} The six terms form three pairs. In the
first, the anchor term $\log(\Ivar\!+\!1)$ sets the entropy
scale, as a $\log$-regularised form of the Holevo lower bound $\Svn\!\ge\!\Ivar$ of
\S\ref{sec:background}. Its partner, $\min(H_A,H_B)\,\Ivar$, weights the mutual
information by the smaller of the regions' summed single-site entropies, which is largest for
balanced bipartitions.

\paragraph{Correlator-kernel sums.} These two terms summarise the distribution of
boundary-correlator magnitudes. The concave kernel $\sum_{E_b}\sqrt{|c|}$ is sensitive
to small correlators, and the saturating one, $\sum_{E_b}\tanh|c|$, weights the larger
correlators. Its $\tanh$ and $\arctan$ versions tie in in-distribution MAE to four decimals
(\S\ref{sec:exhaustive}), so the data fix the saturating shape, which either function
can represent. The concave term is more specific ($\sqrt{|c|}$ has ${\approx}1$~mnat
lower MAE than its cube-root analogue), and we read it as a sub-linear emphasis on
small correlators.

\paragraph{Extremal-correlator terms.} The term $(\max_{E_b}|c|)^2$ is set by the
strongest pair correlation on the boundary, and $\min_{E_b}c$ is the most negative
boundary correlation.

\subsection{Swap and add tests}\label{sec:addswap}

Eq.~\ref{eq:1} is also swap-optimal over the full panel. Swapping any one of its six features
for another panel feature never lowers both in-distribution and OOD MAE, so among single changes, lowering both needs a seventh term. The only stable seventh term lowers MAE by under
$0.5$~mnat; every larger reduction destabilises the slopes. Appending the untransformed
$\Ivar$, for instance, lowers OOD-mean MAE by $1.7$~mnat but raises the slope CV from
$5\%$ to $11\%$, and other additions reach $2$--$19\times$ the CV of Eq.~\ref{eq:1}.
No addition closes the $6.4\times$ gap to the GNN (Table~\ref{tab:in_dist_mae}), so we
keep the six interpretable terms.
\section{Performance}\label{sec:eval}

\subsection{In distribution}\label{sec:indist}

\begin{table}[!ht]\centering\small\renewcommand{\arraystretch}{1.1}
\begin{tabular}{lrccc}
\toprule $N_y$ & $n$ & GNN MAE & Eq.~\ref{eq:1} MAE & $\Ivar$ MAE \\\midrule
1 & 5000 & \textbf{0.0055} & 0.0086 & 0.0515 \\
2 & 5000 & \textbf{0.0044} & 0.0297 & 0.1017 \\
3 & 5000 & \textbf{0.0026} & 0.0180 & 0.0721 \\
4 & 5000 & \textbf{0.0031} & 0.0290 & 0.1095 \\
5 & 5000 & \textbf{0.0032} & 0.0229 & 0.1007 \\
6 & 5000 & \textbf{0.0038} & 0.0384 & 0.1659 \\\midrule
\textbf{All} & 30000 & \textbf{0.0038} & 0.0244 & 0.1002 \\
\bottomrule\end{tabular}
\caption{Per-$N_y$ MAE on the full $30{,}000$-row evaluation set. The paired
GNN-versus-Eq.~\ref{eq:1} gap is $+20.7$~mnat; its paired-bootstrap CI excludes zero
in every row.}\label{tab:in_dist_mae}
\end{table}

With six readable slopes, Eq.~\ref{eq:1} closes $\approx\!79\%$ of the MAE gap
from the classical mutual information ($0.1002$) to the GNN ($0.0038$;
Table~\ref{tab:in_dist_mae}), which remains more accurate. The $25$-term linear fit is
also slightly more accurate than Eq.~\ref{eq:1} (Table~\ref{tab:baselines_panel}), but
its slopes cannot be interpreted, because its $25$-term design matrix has
condition number $\sim\!10^{9}$. Under a training-set bootstrap, $17$ of its $25$ slopes
flip sign, whereas all six slopes of the closed form keep their sign (maximum CV
$5.2\%$).

\subsection{Baselines on the same shortlist}\label{sec:baseline_panel}

To measure what Eq.~\ref{eq:1} gives up in accuracy, we compare it with XGBoost~\cite{ChenGuestrin2016}, a
random forest~\cite{Breiman2001}, a multilayer perceptron (MLP) and the $25$-term linear fit
(Table~\ref{tab:baselines_panel}). They are fit on the same $25$-feature shortlist and
training/evaluation split. On the full evaluation set, the nonlinear baselines have $2$--$3\times$ lower MAE than the
six-feature closed form.

\begin{table}[!ht]\centering\small\renewcommand{\arraystretch}{1.15}
\begin{tabular}{lccccccc}
\toprule predictor & $N_y\!=\!1$ & $2$ & $3$ & $4$ & $5$ & $6$ & \textbf{All}\\\midrule
GNN ($19.9$M)                & 0.0055 & 0.0044 & 0.0026 & 0.0031 & 0.0032 & 0.0038 & \textbf{0.0038}\\
Eq.~\ref{eq:1} (6-feat OLS)  & 0.0086 & 0.0297 & 0.0180 & 0.0290 & 0.0229 & 0.0384 & 0.0244\\
XGBoost (depth-6, 500 trees) & 0.0022 & 0.0063 & 0.0060 & 0.0117 & 0.0102 & 0.0213 & 0.0096\\
Random forest (500 trees)    & 0.0023 & 0.0068 & 0.0065 & 0.0136 & 0.0114 & 0.0239 & 0.0107\\
MLP (256-128-64)             & 0.0044 & 0.0087 & 0.0100 & 0.0136 & 0.0133 & 0.0241 & 0.0124\\
linear, all $25$ (OLS)       & 0.0081 & 0.0239 & 0.0148 & 0.0258 & 0.0210 & 0.0362 & 0.0216\\
$\Ivar$ (Holevo bound, no fit) & 0.0515 & 0.1017 & 0.0721 & 0.1095 & 0.1007 & 0.1659 & 0.1002\\
\bottomrule\end{tabular}
\caption{In-distribution MAE of the predictors fit on the same $25$-feature shortlist,
with the GNN and the classical mutual information for reference.}\label{tab:baselines_panel}
\end{table}

\subsection{Out of distribution}\label{sec:ood}

\begin{table}[!ht]\centering\small\renewcommand{\arraystretch}{1.1}
\begin{tabular}{lrcccc}
\toprule
Pool & $n$ & GNN-base & GNN-ft & Eq.~\ref{eq:1} & $\Ivar$ \\\midrule
eval     & 30000 & \textbf{0.0038} & 0.0226 & 0.0244 & 0.1002 \\
$N_y\!=\!7$  & 1500 & 0.0350 & \textbf{0.0094} & 0.0332 & 0.1486 \\
$N_y\!=\!8$  & 1500 & 0.1127 & \textbf{0.0121} & 0.0597 & 0.2047 \\
$N_y\!=\!9$  & 1500 & 0.1234 & \textbf{0.0433} & 0.0434 & 0.1820 \\
$N_y\!=\!10$ & 1500 & 0.2607 & 0.1044 & \textbf{0.0843} & 0.2470 \\
struct ($3\!\times\!N_y$) & 1500 & 0.0942 & 0.0790 & \textbf{0.0328} & 0.1276 \\
randpos 2-D & 1500 & 0.2402 & 0.2417 & \textbf{0.0334} & 0.1046 \\
\bottomrule\end{tabular}
\caption{MAE on the evaluation set and the six OOD pools. Bold marks the per-pool minimum among the columns shown; GNN-ft and Eq.~\ref{eq:1} tie statistically at $N_y\!=\!9$, as do GNN-base
and Eq.~\ref{eq:1} at $N_y\!=\!7$.}\label{tab:ood_mae}
\end{table}

\paragraph{Comparison with the base GNN.} Out of distribution, Eq.~\ref{eq:1},
fit once and applied unchanged, has $1.9$--$7.2\times$ lower MAE than the base GNN on five OOD pools
($N_y\!\in\!\{8,9,10\}$, struct and randpos). On each of them, the paired-bootstrap CI on $\Delta$MAE, the MAE of Eq.~\ref{eq:1} minus that of the base GNN, excludes zero. On the remaining one, $N_y\!=\!7$, they tie (paired CI
$[-4.2,+0.3]$~mnat, paired-test $p\!=\!0.10$).

\paragraph{Model-per-regime GNN results.} The GNN columns of Table~\ref{tab:ood_mae}
come from different checkpoints, so taking the better in each pool is a
model-per-regime comparison, where each regime may use its own trained model. The
fine-tuned GNN checkpoint has the lowest MAE at its fine-tuning sizes,
$N_y\!\in\!\{7,8\}$ (\S\ref{sec:setup}), and ties Eq.~\ref{eq:1} at $N_y\!=\!9$ (paired CI
$[-3.0,+3.3]$~mnat). Its in-distribution MAE, however, is $5.9\times$ the base GNN's
(from $0.0038$ to $0.0226$), so fine-tuning gained accuracy at the sizes it targeted at the
cost of in-distribution accuracy. Eq.~\ref{eq:1} uses one fit for every pool, so it involves no such trade-off between regimes. On the
geometry-shifted pools neither checkpoint is more accurate than the closed form.

\paragraph{XGBoost out of distribution.} Fit once on the training set as in
Table~\ref{tab:baselines_panel}, XGBoost, unlike the base GNN, has lower MAE than
Eq.~\ref{eq:1} on all six OOD pools, by $8$--$20$~mnat. With all $6{,}000$ training labels and
$N\!\le\!20$, it is the most accurate predictor on these features in distribution and
on every OOD pool except $N_y\!=\!10$, where the random forest has lower MAE.
Table~\ref{tab:label_ladder} gives the ordering at larger sizes with few labels.

\paragraph{Inputs under geometry shift.} Eq.~\ref{eq:1} takes no atomic coordinates
(only $\Ivar$, boundary-correlator summary statistics and single-site entropies, all
computed from bitstrings and the mask). The GNN instead ingests atom
positions and the per-edge angle $\theta_{ij}$ and distance $r_{ij}/\sqrt N$ explicitly
(\S\ref{sec:setup}). The correlator values encode the geometry only implicitly. On the geometry-shifted
pools the GNN's coordinate-dependent map is evaluated off its training geometry,
while the map of Eq.~\ref{eq:1} from correlator structure to entropy is unchanged. The
closed-form margins there ($7.2\times$ on randpos, the largest in
Table~\ref{tab:ood_mae}, and $2.9\times$ on struct) are consistent with this
difference in inputs.
\section{Cross-Hamiltonian transfer}\label{sec:generality}

We next vary the Hamiltonian itself. Refit per family,
Eq.~\ref{eq:1} has a lower MAE than the classical and the filtered
mutual information on every family we tested. XGBoost on the same shortlist is more
accurate but opaque, and the six slopes of the closed form are family-specific.

\subsection{Families and protocol}\label{sec:gen_hams}

We test five families on two-dimensional grids ($3\!\times\!4$, $4\!\times\!4$,
$4\!\times\!5$; $N\!=\!12,16,20$). Throughout, $\langle ij\rangle$ are nearest-neighbour
and $\langle\langle ij\rangle\rangle$ diagonal next-nearest bonds, $\sigma^{x,y,z}_i$ are
the Pauli operators, $\mathbf{S}_i\!=\!\tfrac12\boldsymbol{\sigma}_i$, and
$n_i\!=\!\tfrac12(1\!-\!\sigma^z_i)$ is the occupation; couplings are drawn uniformly over
the stated ranges.
\begin{itemize}\setlength\itemsep{1pt}
\item \textbf{TFIM} (transverse-field Ising model): $H=-J\sum_{\langle ij\rangle}\sigma^z_i\sigma^z_j-h\sum_i\sigma^x_i$;\;
$J\!\in\![0.5,2.0]$, $h\!\in\![0.3,3.0]$.
\item \textbf{XXZ}: $H=\sum_{\langle ij\rangle}\big[J_{xy}(S^x_iS^x_j+S^y_iS^y_j)+J_zS^z_iS^z_j\big]$;\;
$J_{xy}\!\in\![0.5,1.5]$, $J_z\!\in\![0.3,2.5]$.
\item \textbf{$J_1$-$J_2$}: $H=J_1\sum_{\langle ij\rangle}\mathbf{S}_i\!\cdot\!\mathbf{S}_j
+J_2\sum_{\langle\langle ij\rangle\rangle}\mathbf{S}_i\!\cdot\!\mathbf{S}_j$;\;
$J_1\!=\!1$, $J_2\!\in\![0,1.2]$.
\item \textbf{Rydberg array} (the physical system this work targets):
$H=\tfrac{\Omega}{2}\sum_i\sigma^x_i-\Delta\sum_i n_i+\sum_{i<j}V_{ij}\,n_in_j$, with van
der Waals $V_{ij}\!=\!(R_b/a)^6/r_{ij}^6$ over all pairs at the grid distances $r_{ij}$;\;
$\Omega\!=\!1$, $\Delta/\Omega\!\in\![0.5,5.0]$, $R_b/a\!\in\![0.5,3.0]$.
\item \textbf{random-sign} stress test:
$H=\sum_{\langle ij\rangle}J_{ij}\,\mathbf{S}_i\!\cdot\!\mathbf{S}_j$, each
$J_{ij}\!\sim\!\mathcal{N}(0,1)$ drawn independently.
\end{itemize}
In the four \emph{sign-definite} families each bond coupling has a fixed sign, so the
boundary correlators $c_{ij}$ carry a coherent sign pattern. The only sign-resolved feature
of Eq.~\ref{eq:1}, $\min_{E_b}c$, can exploit it; random-sign couplings scramble the signs
and remove it.

\subsection{Controlled comparison and cross-family transfer}\label{sec:gen_indist}

For each family we run one controlled experiment on $1000$ independent ground states
($500/350/150$ parameter draws at $N\!=\!12/16/20$), each with one random bipartition. We
refit the closed form on each family and compare it with XGBoost on the same shortlist,
both fit on identical rows and scored over five $80/20$ splits
(Table~\ref{tab:cross_ham_indist}).

\begin{table}[!ht]\centering\small\renewcommand{\arraystretch}{1.15}
\begin{tabular}{lrccccc}
\toprule
Family & $n_{\text{test}}$ & $\sigma_S$ (mnat) & Eq.~\ref{eq:1} refit & XGBoost (25-feat) & $\Ivar$ & XGB$-$Eq.~\ref{eq:1} (mnat)\\\midrule
TFIM             & 200 & $196$ & 0.0285 & \textbf{0.0102} & 0.110 & $-18.3$ \\
XXZ              & 200 & $639$ & 0.0466 & \textbf{0.0216} & 0.130 & $-25.0$ \\
$J_1$-$J_2$      & 200 & $849$ & 0.0695 & \textbf{0.0386} & 0.277 & $-30.9$ \\
Rydberg          & 200 & $354$ & 0.0374 & \textbf{0.0214} & 0.150 & $-15.9$ \\
random-sign      & 200 & $792$ & 0.0619 & \textbf{0.0351} & 0.130 & $-26.7$ \\
\bottomrule\end{tabular}
\caption{Controlled cross-Hamiltonian comparison on 2-D grids. For each family the three
grids ($N\!=\!12,16,20$) are combined into one $1000$-row dataset, with one bipartition
per state. Eq.~\ref{eq:1} and XGBoost are fit on these rows. All four predictors
(Eq.~\ref{eq:1}, XGBoost, $\Ivar$ and $\Ivar_{\mathrm{filt}}$ of
Fig.~\ref{fig:headline}b) are scored on them. Each MAE, in nats, is the mean over five $80/20$ splits
($n_{\text{test}}\!=\!200$). $\sigma_S$ is the per-family standard deviation of the target
$\Svn$, a measure of target diversity. Eq.~\ref{eq:1}
has a lower MAE than $\Ivar$ and than $\Ivar_{\mathrm{filt}}$ (Fig.~\ref{fig:headline}b; values in SI Table~\ref{tab:crossham_ext}) on every family. XGBoost has a lower MAE than Eq.~\ref{eq:1}
by $15.9$--$30.9$~mnat, the same ordering as on the evaluation set
(Table~\ref{tab:baselines_panel}).}\label{tab:cross_ham_indist}
\end{table}

XGBoost is more accurate than Eq.~\ref{eq:1} in every family, but on the four sign-definite families the MAE of the closed form is $2.8$--$4.0\times$ lower than that of $\Ivar$.
Its pooled MAE there is $29$--$70$~mnat, and in most families its per-grid MAE rises with
grid size (SI Table~\ref{tab:crossham_ext}), consistent with the scale effect of
\S\ref{sec:residuals}. On random-sign the margin over $\Ivar$ falls to $2.1\times$, the
smallest of any family, consistent with $\min_{E_b}c$ being most useful where the correlators carry a coherent sign.

The table has no per-family GNN column, because retraining the GNN per family is out
of scope.
We would expect the GNN to stay more accurate than Eq.~\ref{eq:1} and XGBoost here, as on
the evaluation set (Table~\ref{tab:baselines_panel}). This expectation rests on the
thesis version of the GNN paper~\cite{SalehThesis}, where the same architecture reaches
$2.6$~mnat on transverse-field Ising configurations.

To test whether the fits themselves transfer, we fit each predictor on family $X$ and evaluate it zero-shot on
family $Y$, with no refit on $Y$ (full $5\!\times\!5$ table in SI Table~\ref{tab:cross_ham_transfer}). On the diagonal (same family) the MAE stays at tens
of mnat. Off the diagonal the MAE is larger, by one to
two orders of magnitude in the worst cells, and several cells exceed $1$~nat. The six
features therefore carry over between families when Eq.~\ref{eq:1} is refit, and the slopes
do not. Comparing the predictors, neither transfers uniformly better. Eq.~\ref{eq:1} has the
lower MAE in $9$ of the $20$ off-diagonal cells and XGBoost in $10$, with $1$ tie.

\section{Residuals, size scaling and shot noise}\label{sec:diag}

\subsection{Residuals}\label{sec:residuals}

\begin{figure}[!ht]\centering
\includegraphics[width=0.95\textwidth]{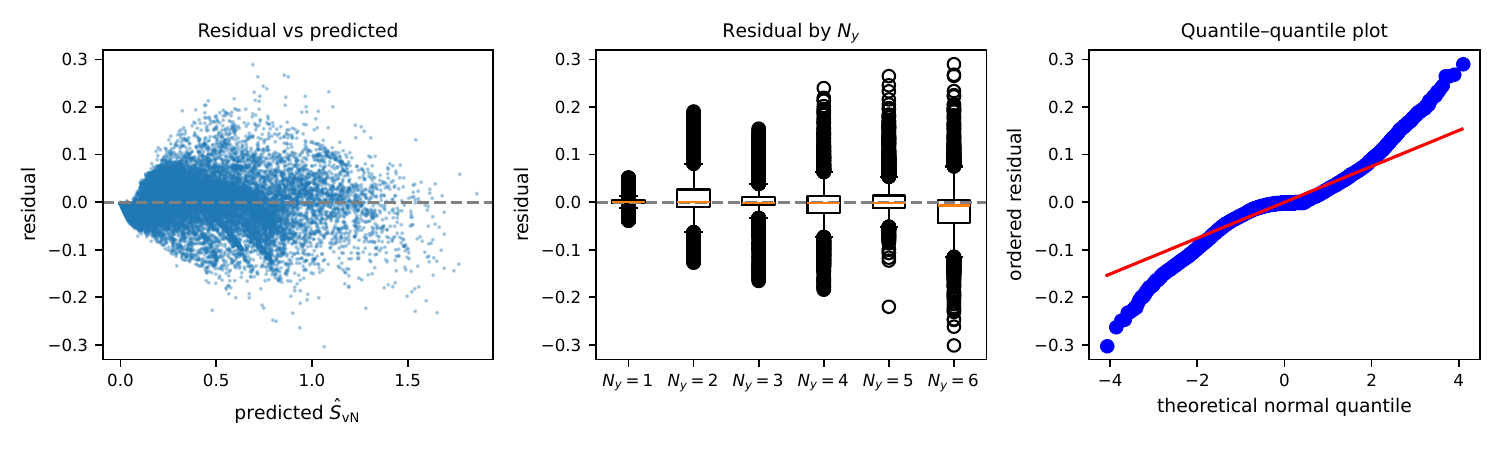}
\caption{Eq.~\ref{eq:1} residuals on the evaluation set. \emph{Left:} residual against the
predicted entropy $\hat\Svn$; the spread widens with the prediction.
\emph{Middle:} residual by ladder length $N_y$; the interquartile range and the tails grow
overall with $N_y$. \emph{Right:} quantile--quantile plot, which shows the heavy tails. The
dispersion concentrates at higher entropy, consistent with
$\mathrm{corr}(|r|,\Svn)\!=\!0.47$.}\label{fig:residuals}
\end{figure}

On the evaluation set, the residuals $r$ of Eq.~\ref{eq:1} (true minus predicted
entropy) are heteroscedastic and heavy-tailed (Fig.~\ref{fig:residuals}). Their mean at each $N_y$ varies
non-monotonically between $+6$ and $-14$~mnat and is most negative at $N_y\!=\!6$. The
residuals are smallest for near-product states and grow with the true entropy
($\mathrm{corr}(|r|,\Svn)\!=\!0.47$; the $99$th percentile rises from $0.02$ to $0.17$~nats
between $\Svn\!<\!0.01$ and $\Svn\!>\!0.6$, SI~\S\ref{app:worst}). This growth is largely a scale effect, since the
\emph{relative} error $|r|/\Svn$ falls from $\sim\!16\%$ at $\Svn\!\approx\!0.2$ to $\sim\!5\%$
for $\Svn\!>\!0.6$ ($\mathrm{corr}(|r|/\Svn,\Svn)\!=\!-0.42$; a linear fit gives
$|r|\!\approx\!0.044\,\Svn+0.010$). The closed form is thus most accurate in relative terms on
the most entangled states.

We then ask whether the worst-fit states are hard. On the $50$ rows with the
largest residuals, the GNN is far more accurate than
Eq.~\ref{eq:1} (median $6.5$~mnat, all but one below $20$, against $190$--$300$~mnat; SI~\S\ref{app:worst}).
These states are not hard for the GNN, so we attribute the large closed-form residuals to
the limited capacity of six features.

Residual size is also partly predictable from the features alone. A linear model on the
same six features predicts $|r|$ out-of-fold with a Spearman correlation of $0.77$, so as a
trust score it would rank rows by expected residual size.

\subsection{Size scaling}\label{sec:size}

The slopes of Eq.~\ref{eq:1} are not scale-invariant. To see how they vary, we refit them
separately for each even ladder length (Table~\ref{tab:size_scaling}). From $N\!=\!4$ to $N\!=\!20$ the anchor slope $b_1$ (the
weight on $\log(\Ivar\!+\!1)$) roughly doubles, while the magnitude $|b_5|$ of
the $\min_{E_b}c$ weight falls towards zero. All six slopes vary with $N$ over this range. Accordingly,
the MAE of the global fit, applied unchanged, rises from $0.024$~nats in distribution to $0.084$~nats
at $N\!=\!20$. A per-size refit recovers ${\approx}30$~mnat of that error, so the large-$N$
growth comes partly from slope drift and partly from the larger entropy scale of
\S\ref{sec:residuals}.

We expect the boundary-restricted feature \emph{set} to remain the right basis at larger
$N$, but its slopes depend on size. Reaching arrays of about $100$ atoms therefore needs either a
per-size refit or an explicit slope-versus-$N$ model in place of the fixed slopes.
Our ED pipeline stops at $N\!=\!20$, so \S\ref{sec:dmrg} tests both
options against DMRG datasets up to $N\!=\!100$. The bipartition ensembles differ, however.
Table~\ref{tab:size_scaling} refits on the random masks of the evaluation set and
the size-OOD pools, whereas \S\ref{sec:dmrg} uses the balanced half cut throughout. The
slope values need not coincide between Table~\ref{tab:size_scaling} and Fig.~\ref{fig:bnfits}, so only their trends in $N$ are comparable.

\begin{table}[!ht]\centering\small\renewcommand{\arraystretch}{1.1}
\begin{tabular}{rrrrrrrcc}
\toprule
$N$ & $b_1$ & $b_2$ & $b_3$ & $b_4$ & $b_5$ & $b_6$ & MAE$_{\mathrm{refit}}$ & MAE$_{\mathrm{global}}$\\\midrule
$4$  & $0.98$ & $0.22$ & $-0.55$ & $-4.26$ & $-1.58$ & $0.23$ & $0.026$ & $0.030$\\
$8$  & $1.25$ & $0.15$ & $-0.38$ & $-0.92$ & $-0.37$ & $0.13$ & $0.026$ & $0.029$\\
$12$ & $1.59$ & $0.09$ & $-0.28$ & $-4.31$ & $-0.50$ & $0.10$ & $0.033$ & $0.038$\\
$16$ & $1.70$ & $0.06$ & $-0.16$ & $-5.37$ & $-0.33$ & $0.05$ & $0.047$ & $0.060$\\
$20$ & $2.07$ & $0.04$ & $-0.13$ & $-6.72$ & $-0.16$ & $0.04$ & $0.053$ & $0.084$\\
\bottomrule\end{tabular}
\caption{Eq.~\ref{eq:1} slopes refit separately for each even-length $2\!\times\!N_y$ ladder
($N\!=\!2N_y$). The MAE columns give, at each size, the MAE of this refit and that of
the global Eq.~\ref{eq:1} applied unchanged.}\label{tab:size_scaling}
\end{table}

\subsection{Shot noise}\label{sec:shot}

To model finite measurement statistics, we draw $S$ shots from $|\psi|^2$ and recompute the
six features from these shots. We sweep $S$ on a random $5{,}000$-row sample of the
evaluation set and score Eq.~\ref{eq:1}, $\Ivar$ and the
$25$-feature XGBoost, with every input recomputed from the same shots.

On the exact distribution, Eq.~\ref{eq:1} and $\Ivar$ have MAEs of $0.024$ and
$0.100$~nats, as in \S\ref{sec:indist}. The MAE curves
of the closed form and $\Ivar$ cross at $S\!\approx\!160$ (interpolated between
sweep points). At ${\sim}150$ shots and below the empirical correlators are
too noisy, so $\Ivar$ is comparable or better. Above the
crossover the closed form has the lower MAE, by a factor that grows to $1.4\times$ by
$S\!=\!500$ and to $4.2\times$ on the exact distribution. Its MAE falls below $0.05$~nats
near $S\!\approx\!1430$.

\begin{figure}[!ht]\centering
\includegraphics[width=0.68\textwidth]{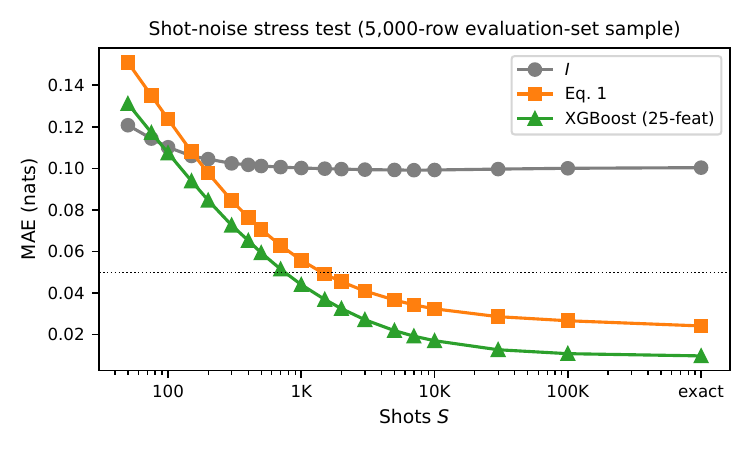}
\caption{MAE versus the number of measurement
shots $S$ for Eq.~\ref{eq:1}, XGBoost, and the classical mutual information $\Ivar$. The dotted line marks $0.05$~nats.}\label{fig:shotnoise}
\end{figure}

XGBoost is the most accurate for
$S\!\ge\!100$, and from $S\!=\!150$ up its MAE is ${\sim}11$--$15$~mnat below that of
Eq.~\ref{eq:1} (Fig.~\ref{fig:shotnoise}). Below $S\!\approx\!100$ the noisy correlators
degrade XGBoost too, and $\Ivar$ has the lower MAE.

\section{Validation at scale: DMRG to \texorpdfstring{$N=100$}{N=100}}\label{sec:dmrg}

Every result so far rests on ED labels, which end at $N\!=\!20$. This section therefore
asks the deployment question, whether slopes fitted at $N\!\le\!12$ extrapolate to larger sizes. To
answer it, we compute ground-state entropies up to $N\!=\!100$, five times our ED reach, with DMRG.

\subsection{Design}\label{sec:dmrg_design}

\paragraph{Setup.} We solve $2\!\times\!N_y$ Rydberg ladders with the same Hamiltonian conventions
as \S\ref{sec:setup}. The solver is two-site DMRG~\cite{White1992,Schollwoeck2011} as implemented in
TeNPy~\cite{TenPy}. The ladder lengths are
$N_y\!\in\!\{12,\allowbreak14,\allowbreak16,\allowbreak18,\allowbreak20,\allowbreak22,\allowbreak24,\allowbreak26,\allowbreak28,\allowbreak30,\allowbreak34,\allowbreak38,\allowbreak42,\allowbreak46,\allowbreak50\}$, i.e.\
fifteen sizes spanning $N\!=\!24$--$100$, with $100$ convergence-certified rows per size. Parameters
are drawn from the OOD convention ($\Delta/\Omega\!\sim\!U[0,6]$, $R_b/a\!\sim\!U[0.1,5]$). The bond
dimension is raised through $\chi\!=\!128$, $256$, then $512$ until the cut entropy converges; the
convergence checks that certify each row are given in SI~\S\ref{si:dmrg}. The bipartition is the
balanced half cut, in which subsystem $A$ is the first $N_y/2$ rungs. Because this contiguous
cut differs from the random masks of the main pools, an ED reference band at $N\!=\!8$--$20$ uses the
same half cut and parameter distribution, with $200$ rows per size. It shares its sizes with the
random-mask pools and its cut with the DMRG rows, so it separates the mask shift from the size shift. A small-size band of half-cut labels at
$N\!\le\!12$, computed with the same DMRG pipeline, serves to pretrain the frozen and fine-tuned six-feature XGBoost comparators of \S\ref{sec:dmrg_results}.
For every DMRG row, the occupations and boundary correlators are computed exactly from the
matrix-product state. The classical mutual information is estimated from $20{,}000$
configurations drawn from the state by perfect sampling. Each of its three Shannon entropies is the
mean of $-\log p$ under its own distribution $p$: the joint distribution of all sites for
$H(X_A,X_B)$ and the marginal on $A$ or $B$ for the other two. We estimate each one as the average of
$-\log p$ over the samples, with every probability computed exactly from the state. The
estimate is therefore unbiased and differs from the exact value only by sampling noise.

\subsection{Results}\label{sec:dmrg_results}

\begin{table}[!ht]\centering\small\renewcommand{\arraystretch}{1.1}
\begin{tabular}{llrrc}
\toprule
$N$ & labels & $n$ & frozen Eq.~\ref{eq:1} (nats) & refit, $5$-fold CV (mnat, $95\%$ CI)\\\midrule
$8$ & ED & 200 & 0.023 & 15.7 $[13,18]$ \\
$10$ & ED & 200 & 0.017 & 11.2 $[10,13]$ \\
$12$ & ED & 200 & 0.065 & 25.7 $[22,29]$ \\
$14$ & ED & 200 & 0.047 & 14.7 $[12,17]$ \\
$16$ & ED & 200 & 0.106 & 25.1 $[22,29]$ \\
$18$ & ED & 200 & 0.081 & 14.9 $[13,17]$ \\
$20$ & ED & 200 & 0.218 & 29.1 $[25,33]$ \\
\midrule
$24$ & DMRG & 100 & 0.252 & 18.2 $[15,22]$ \\
$28$ & DMRG & 100 & 0.379 & 31.0 $[23,40]$ \\
$32$ & DMRG & 100 & 0.367 & 31.8 $[26,38]$ \\
$36$ & DMRG & 100 & 0.541 & 25.5 $[21,30]$ \\
$40$ & DMRG & 100 & 0.631 & 34.6 $[27,45]$ \\
$44$ & DMRG & 100 & 0.659 & 36.9 $[30,45]$ \\
$48$ & DMRG & 100 & 0.966 & 30.7 $[25,37]$ \\
$52$ & DMRG & 100 & 0.668 & 27.4 $[20,35]$ \\
$56$ & DMRG & 100 & 1.054 & 32.1 $[23,43]$ \\
$60$ & DMRG & 100 & 1.255 & 22.1 $[18,27]$ \\
$68$ & DMRG & 100 & 0.744 & 29.3 $[20,43]$ \\
$76$ & DMRG & 100 & 0.777 & 20.2 $[13,30]$ \\
$84$ & DMRG & 100 & 1.871 & 23.4 $[18,29]$ \\
$92$ & DMRG & 100 & 1.501 & 15.5 $[12,20]$ \\
$100$ & DMRG & 100 & 1.685 & 18.1 $[11,26]$ \\
\bottomrule
\end{tabular}

\caption{Size scaling on the balanced half cut. For each size the table gives the label source, the
row count $n$ ($100$ DMRG rows, $200$ ED reference rows), the MAE of the
frozen global Eq.~\ref{eq:1}, and the per-size-refit MAE with its $5$-fold
cross-validated $95\%$ confidence interval. The MAE of the
frozen fit rises from $0.017$--$0.218$~nats over the ED reference band to
$0.25$--$1.87$~nats at $N\!=\!24$--$100$. The per-size refit stays at
$15$--$37$~mnat at the DMRG sizes ($11$--$29$ over the ED reference band).}\label{tab:dmrg_scaling}
\end{table}

\textbf{(a) The frozen global fit.} Applied unchanged, this fit already has an MAE
of $17$--$218$~mnat over the ED reference band. At $N\!=\!52$--$100$ its MAE rises unevenly to
$0.67$--$1.87$~nats (Table~\ref{tab:dmrg_scaling}), consistent with the slope drift of
\S\ref{sec:size}. Part of the error at the ED sizes is mask shift, since at $N\!=\!20$ the frozen
fit scores $0.084$~nats on the random-mask pool (Table~\ref{tab:size_scaling}) but $0.218$~nats on the balanced
half cut. The further growth to $1.68$~nats at $N\!=\!100$ is the size shift proper. By contrast, a frozen
XGBoost with the same six features, pretrained on the small-size band (\S\ref{sec:dmrg_design}), degrades
only to $34$--$50$~mnat at the larger sizes (Table~\ref{tab:label_ladder}). This points to
the fixed linear slopes as the source of the blow-up, which the slope laws below correct.

\textbf{(b) The per-size refit.} When the slopes are refit at each size instead, the same six features reach
$15$--$37$~mnat at every DMRG size ($11$--$29$ over the ED reference band). These MAEs are 5-fold
cross-validated over all rows of each size. The six-feature design matrix also stays well-conditioned
at scale (condition number $28$--$100$ across all twenty-two sizes, against $\sim\!10^{9}$ for the
$25$-term linear fit of \S\ref{sec:indist}).

\begin{figure}[!ht]\centering
\includegraphics[width=0.98\textwidth]{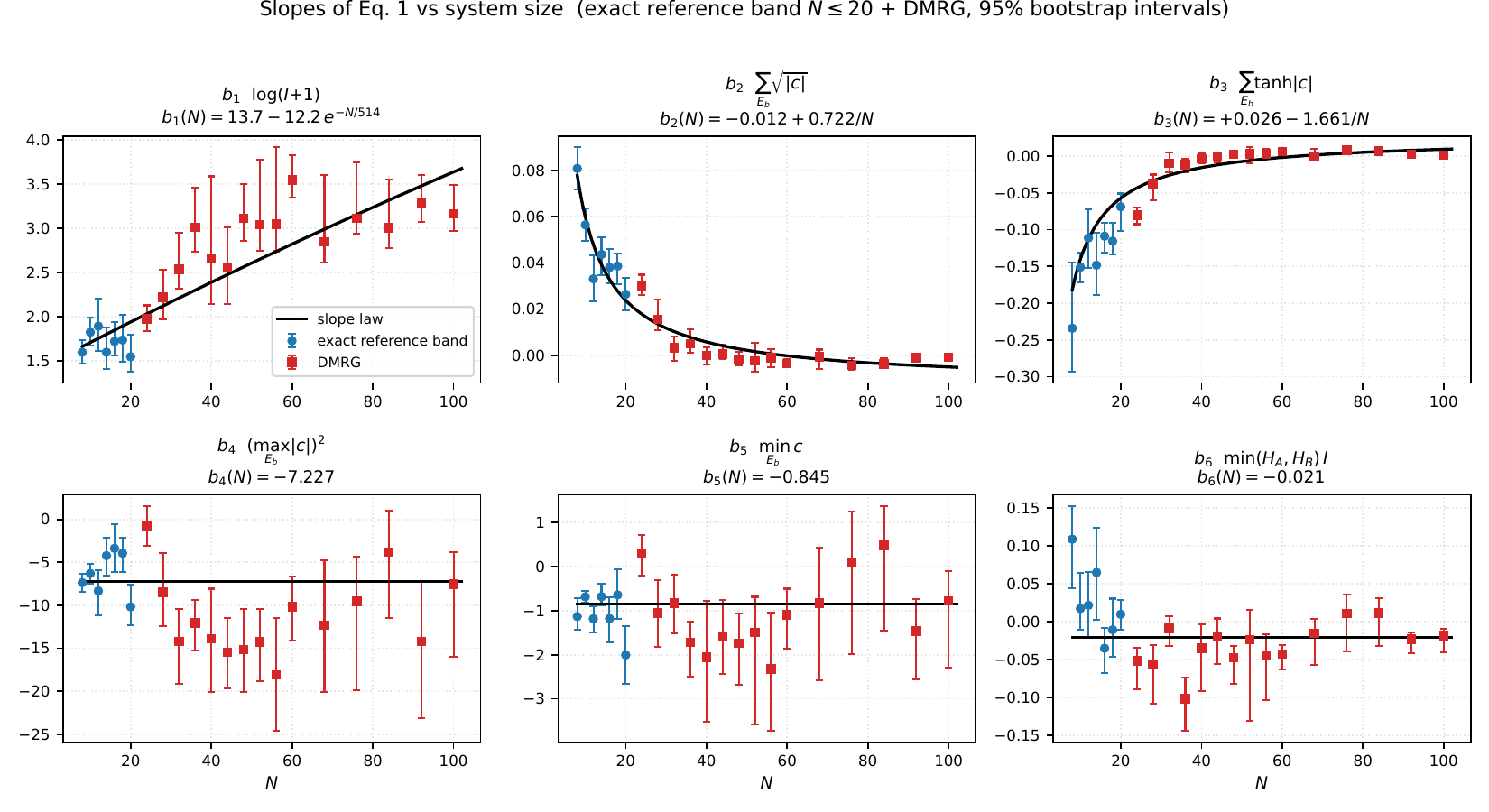}
\caption{Slopes of Eq.~\ref{eq:1} refit per size on the balanced half cut, for the exact
reference band ($N\!=\!8$--$20$, circles)
and DMRG ($N\!=\!24$--$100$, squares, fifteen sizes), with $95\%$
bootstrap intervals. The solid curve in each panel is the law selected by the corrected Akaike information criterion (AICc), with its fitted equation shown. The anchor slope $b_1$ grows and curves downward,
the two correlator-kernel slopes $b_2,b_3$ follow $1/N$ laws, and $b_4,b_5,b_6$ show no resolvable
trend and are held constant.}\label{fig:bnfits}
\end{figure}

\textbf{(c) Slope laws.} To follow the slopes across sizes, Fig.~\ref{fig:bnfits} shows each per-size slope with bootstrap
intervals and the fitted $b_i(N)$ laws (at most two free parameters each; the
saturating-exponential scale is held fixed, SI~\S\ref{si:dmrg}). The anchor slope $b_1$ (on
$\log(\Ivar\!+\!1)$) grows from $\approx\!1.5$--$1.9$ over the exact reference band to
$\approx\!2.8$--$3.5$ at $N\!=\!52$--$100$. The corrected Akaike information criterion (AICc)~\cite{HurvichTsai1989} selects a
saturating exponential for it (weighted $R^2\!=\!0.81$). We keep it rather than logarithmic growth, which fits the per-size slopes comparably ($\Delta$AICc
$\approx\!1$) but degrades the parametrised closed form of (d) to $24$--$138$~mnat over the DMRG
sizes. The exponential's fitted asymptote is $\approx\!14$, but with the scale fixed at $514$ sites the slope would
approach it only far beyond $N\!=\!100$, so no saturation is resolved in the window. We therefore
report $b_1$ as growth that curves downward and do not read the fitted asymptote, the constant in Table~\ref{tab:bN_laws}, as a limiting value. For the two correlator-kernel slopes, $b_2$ ($\sum\!\sqrt{|c|}$)
and $b_3$ ($\sum\tanh|c|$), the AICc-selected law is a $1/N$ finite-size correction (weighted
$R^2\!=\!0.89$ and $0.74$). The remaining slopes $b_4$, $b_5$,
and $b_6$ show no resolvable trend (no law fit reaches weighted $R^2\!>\!0.5$), so we treat them as
constants. Of these, $b_6$ is small throughout ($|b_6|\!\lesssim\!0.1$).

\textbf{(d) The parametrised closed form Eq.~\ref{eq:1}{\boldmath$(N)$}.} Substituting the
fitted laws (Table~\ref{tab:bN_laws}) for the six slopes gives a closed form whose slopes are set by $N$ alone. The laws are fit on all rows,
which also fix the per-size slopes, so the evaluated size itself enters the fit. Evaluated in this way, inside the window, it reaches $16$--$45$~mnat over the exact reference band and
$24$--$98$~mnat at the DMRG sizes. This in-window result is interpolation and uses labels at the target size.
The no-label test refits each law without the target size and evaluates it there (SI Table~\ref{tab:si_loso}).
It gives $15$--$46$~mnat over the exact reference band and $24$--$95$~mnat at the DMRG sizes, close to the in-window values.

\begin{table}[!ht]\centering\small\renewcommand{\arraystretch}{1.1}
\begin{tabular}{cllcl}
\toprule
slope & feature & fitted law $b_i(N)$ & wR$^2$ & $N$-trend \\\midrule
$b_1$ & $\log(\Ivar\!+\!1)$    & $13.7 - 12.2\,e^{-N/514}$          & $0.81$ & grows, downward-curving \\
$b_2$ & $\sum_{E_b}\sqrt{|c|}$  & $-0.012 + 0.72\,N^{-1}$           & $0.89$ & $1/N$ decay \\
$b_3$ & $\sum_{E_b}\tanh|c|$    & $\phantom{-}0.026 - 1.66\,N^{-1}$ & $0.74$ & $1/N$ decay \\
$b_4$ & $(\max_{E_b}|c|)^2$    & $-7.23$ \ (const)                 & $<0.5$ & no resolved trend \\
$b_5$ & $\min_{E_b}c$          & $-0.85$ \ (const)                 & $<0.5$ & no resolved trend \\
$b_6$ & $\min(H_A,H_B)\,\Ivar$ & $-0.021$ \ (const)                & $<0.5$ & no resolved trend \\
\bottomrule
\end{tabular}

\caption{Size-scaling laws for the six slopes of Eq.~\ref{eq:1}, with their weighted $R^2$. The
slopes are refit per size on the balanced half cut.}\label{tab:bN_laws}
\end{table}

\begin{table}[!ht]\centering\small\renewcommand{\arraystretch}{1.1}
\begin{tabular}{lrrrr}
\toprule
labels / estimator & $N\!=\!52$ & $60$ & $84$ & $100$\\\midrule
$0$ \ slope laws Eq.~\ref{eq:1}$(N)$, size left out & 38 & 47 & 77 & 95 \\
$0$ \ frozen XGBoost (small-size pretrain) & 34 & 46 & 50 & 39 \\
$12$ \ refit & 82 & 46 & 52 & 115 \\
$20$ \ refit & 35 & 29 & 29 & 32 \\
$30$ \ refit & 29 & 25 & 25 & 24 \\
$50$ \ refit & 26 & 23 & 24 & 20 \\
\midrule
$20$ \ XGBoost, scratch & 56 & 70 & 59 & 54 \\
$20$ \ XGBoost, fine-tuned & 34 & 47 & 44 & 37 \\
$50$ \ XGBoost, scratch & 33 & 37 & 34 & 32 \\
$50$ \ XGBoost, fine-tuned & 26 & 37 & 33 & 29 \\
\midrule
all \ per-size refit & 27 & 22 & 23 & 18 \\
\bottomrule
\end{tabular}

\caption{Label budget at the target size (MAE, mnat). Each $k$-label entry is the median over $2{,}000$
random draws of $k$ labelled rows, fit on the $k$
labels and scored on the held-out remainder. The $0$-label rows use no labels at the target size. They evaluate the slope
laws of Fig.~\ref{fig:bnfits}, refit without the target size (leave-one-size-out, SI Table~\ref{tab:si_loso}), and the frozen XGBoost pretrained on the small-size band. In the
``XGBoost, scratch'' rows the model is trained on the sampled labels alone; in the
``XGBoost, fine-tuned'' rows it is pretrained on the small-size band and then warm-started on the same labels. The last row is the cross-validated per-size refit of Table~\ref{tab:dmrg_scaling}.}\label{tab:label_ladder}
\end{table}

\textbf{(e) Label budget at the target size.} Table~\ref{tab:label_ladder} gives the MAE at the
target size as a function of the number of labels available there. With no labels, the parametrised
closed form (laws refit without that size) is on par with the frozen XGBoost of (a) at
$N\!=\!52$--$60$ and has higher MAE at the largest sizes. With labels, all six slopes can be refit
at the target size. This refit overfits with ${\sim}12$ labels, reaches $29$--$35$~mnat with
$\approx\!20$, and with $\approx\!50$ reaches the per-size-refit band at these sizes
($18$--$27$~mnat). We compare it with XGBoost at budgets of $20$ and $50$ labels. At both budgets the
refit closed form has lower MAE than XGBoost trained from scratch on the same labels. The stronger
comparator at these budgets is a fine-tuned XGBoost, pretrained on the small-size band and
warm-started on the same labels and six features. Against it, the refit closed form is comparable
at $N\!=\!52$ and has lower MAE at $N\!=\!60$--$100$ at both budgets.

\section{Is the closed form a distillation of the GNN?}\label{sec:decod}

We used the GNN's boundary routing (\S\ref{sec:gnn_boundary}) to set the boundary
restriction of the features.
This section asks the converse question: is Eq.~\ref{eq:1} a distillation of the network, and
could one instead read a closed form directly off it? The GNN's prediction
aligns with physical quantities that Eq.~\ref{eq:1} also uses
(\S\ref{sec:gradsvd}), but it encodes them redundantly, spread over many correlated directions
(\S\ref{sec:dir_ablation}). Reading a closed form off the network yields
one of two readouts (\S\ref{sec:gnn_derived}). A linear readout of the $1024$-dimensional
post-readout state, the output of the network's readout projection, is opaque. A sparse
readout built on sparse-autoencoder (SAE) codes~\cite{Cunningham2024SAE} is less accurate at code counts small enough to read.

\subsection{Held-out decodability}\label{sec:probe_decod}

\begin{figure}[h]\centering
\includegraphics[width=0.82\textwidth]{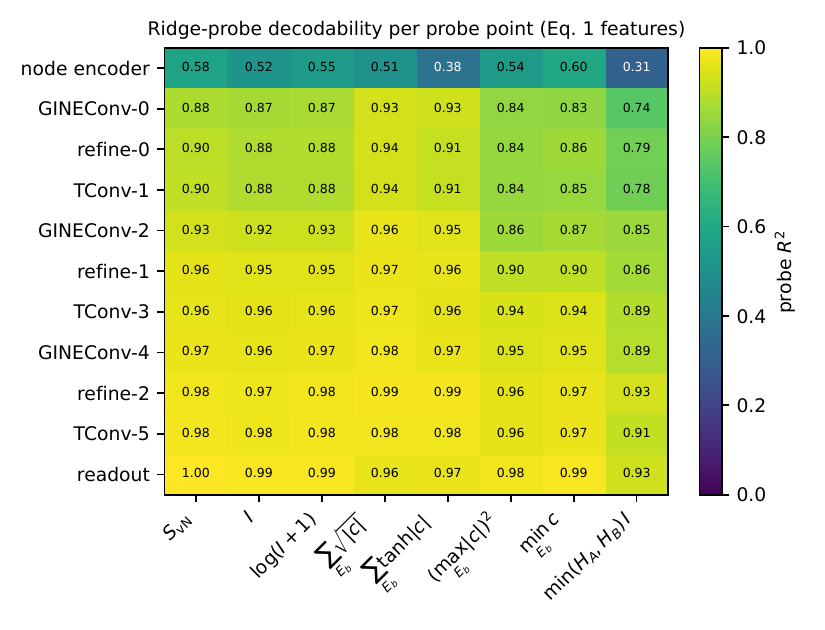}
\caption{Held-out ridge-probe $R^2$ at each probe point for $\Svn$, $\Ivar$, and the six
features. Probe points run from input (top) to output (bottom): the node encoder, the six
message-passing layers of \S\ref{sec:setup}, the per-node refinement blocks that follow
each GINEConv layer in the published architecture, and the readout projection. At the readout,
$R^2\!\ge\!0.996$ for $\Svn$ and $R^2\!\ge\!0.92$ for every feature, although two of the
boundary correlator sums peak at earlier probe points.}\label{fig:decod}
\end{figure}

We start by testing whether the network carries these features. A ridge-regression probe recovers every Eq.~\ref{eq:1} feature from the post-readout state at
$R^2\!\ge\!0.92$, and $\Svn$ itself at $R^2\!=\!0.996$ (Fig.~\ref{fig:decod}). The probe uses a
13-point $\lambda$-sweep with 5-fold cross-validation for $\lambda$ and is evaluated on a held-out
$30\%$ split.
$R^2$ generally rises through the message-passing layers, although some features dip slightly at individual probe points. But decodability establishes
only that a quantity is linearly present, not how the prediction uses it;
\S\ref{sec:gradsvd} examines that use, and \S\ref{sec:dir_ablation} shows that the
quantities are encoded redundantly.

\subsection{The information-theoretic and extremal-correlator terms organise the post-readout state}\label{sec:embed}

\begin{figure}[h]\centering
\includegraphics[width=0.95\textwidth]{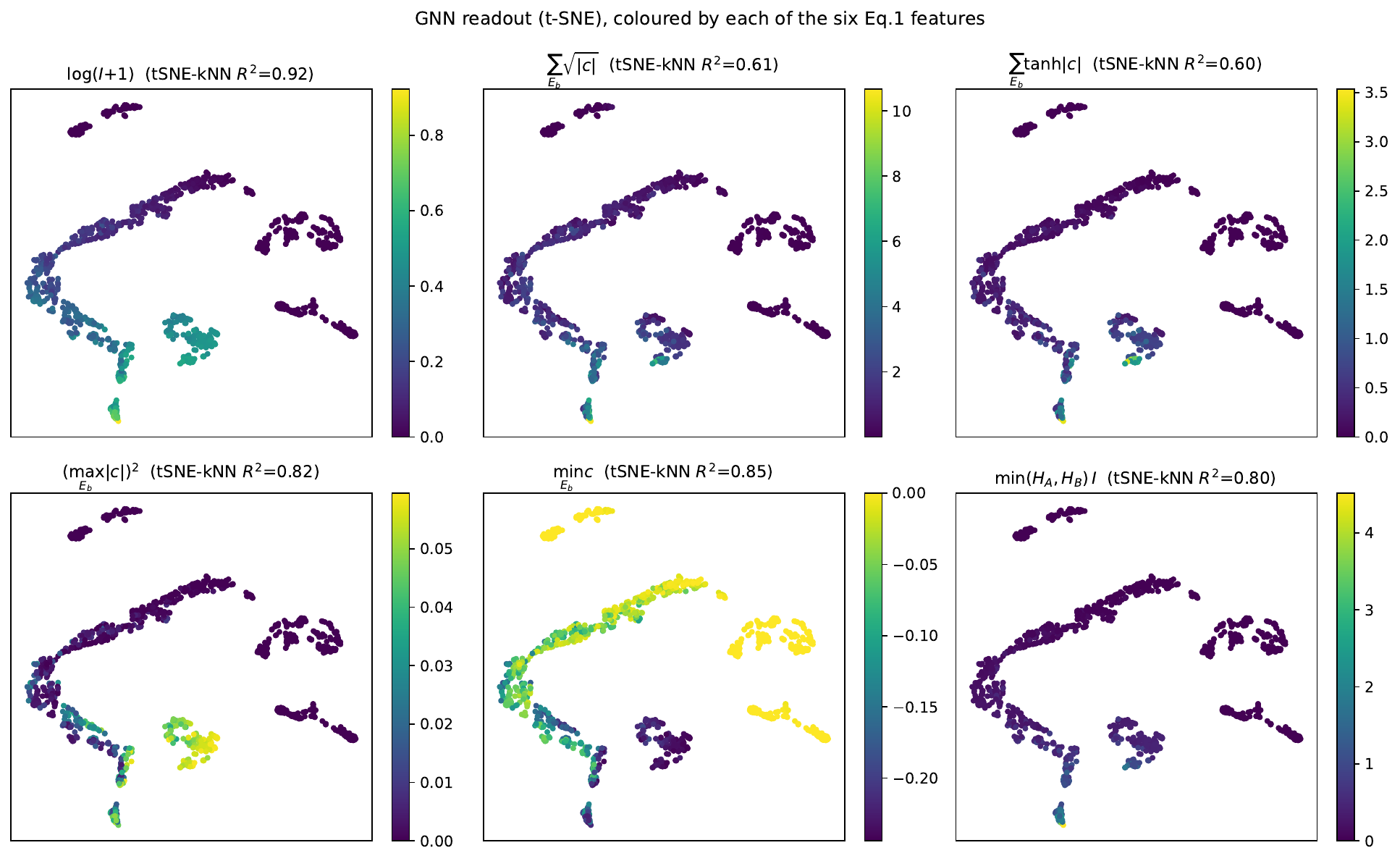}
\caption{$2$-D t-SNE of the GNN post-readout state ($1{,}200$ graphs), coloured by each
of the six Eq.~\ref{eq:1} features (cross-validated kNN $R^2$ in each title). The
embedding separates by the information-theoretic pair and the extremal-correlator
terms ($R^2\!=\!0.80$--$0.92$) and more weakly by the two correlator-kernel sums
($\approx\!0.60$).}\label{fig:embed}
\end{figure}

The Eq.~\ref{eq:1} features also organise the post-readout state. A principal-component analysis (PCA) of the post-readout state puts
$62.6\%$ of its variance in the top two components and $97.5\%$ in the top ten, and
those components track the features. We score each quantity by the cross-validated
k-nearest-neighbour (kNN) $R^2$ for predicting it from the $2$-D coordinates of a
t-distributed stochastic neighbour embedding (t-SNE)~\cite{vanderMaaten2008}. Values from the top-$2$\,/\,top-$10$ PCs follow in parentheses. The target $S_{\mathrm{vN}}$ reaches $0.96$
($0.74$/$0.96$). For the six Eq.~\ref{eq:1} features they are $0.92$ for
$\log(\Ivar+1)$ ($0.76$/$0.93$), $0.85$ for $\min_{E_b}c$ ($0.65$/$0.83$), $0.82$ for
$(\max_{E_b}|c|)^2$ ($0.51$/$0.79$), and $0.80$ for $\min(H_A,H_B)\,\Ivar$ ($0.31$/$0.79$).
The two correlator-kernel sums, $\sum_{E_b}\!\sqrt{|c|}$ and $\sum_{E_b}\!\tanh|c|$,
are more weakly organised, at $0.61$ and $0.60$ ($0.20$/$0.53$ and $0.28$/$0.48$)
(Fig.~\ref{fig:embed}). The information-theoretic pair and the extremal-correlator
terms therefore organise the post-readout state most strongly.

\subsection{The prediction's readout gradients are low-dimensional}\label{sec:gradsvd}

\begin{figure}[!ht]\centering
\includegraphics[width=0.95\textwidth]{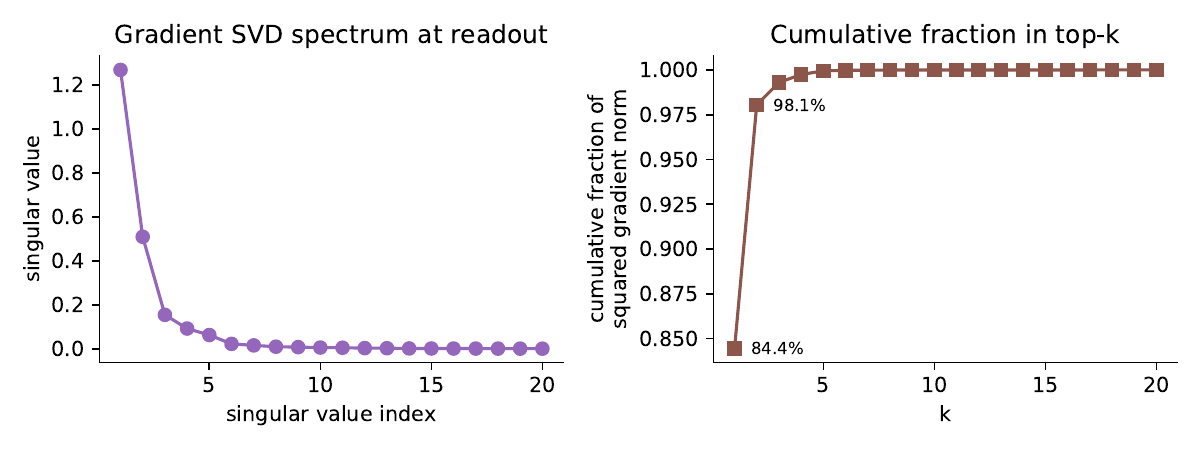}
\caption{SVD of the readout gradients, computed on $n\!=\!2{,}200$ in-distribution
graphs. \emph{Left:} singular-value spectrum of the per-graph readout gradients
$\partial\hat\Svn/\partial h^{\mathrm{readout}}$, which drops steeply after the first
one or two. \emph{Right:} cumulative fraction of the squared gradient norm in the top
$k$ directions: $84.4\%$ in the first, $98.1\%$ in the top two, $99.9\%$ in the top
five.}\label{fig:gradsvd}
\end{figure}

To see which directions the prediction uses, we compute per-graph gradients of the prediction with respect to the post-readout
state, $\partial\hat\Svn/\partial h^{\mathrm{readout}}$, on $2{,}200$ in-distribution
graphs. We then take the singular-value decomposition (SVD) of the resulting matrix. The top
right singular vector carries $84.4\%$ of the (uncentred) squared gradient norm, the
top two carry $98.1\%$, and the top five $99.9\%$ (Fig.~\ref{fig:gradsvd}). In correlations with the candidate panel of \S\ref{sec:panel}, the leading one tracks $\Ivar$, the second
the boundary-correlator sign structure, the third $(\max_{E_b}|c|)^2$, and the fourth the
mean density $\bar n$.

\subsection{Direction ablation and redundant encoding}\label{sec:dir_ablation}

Projecting the span of the (held-out) Eq.~\ref{eq:1} feature probes out of the readout,
and keeping its complement, tests whether the prediction needs these directions. It raises the MAE by $2.4$~mnat over the GNN's $\sim\!4$~mnat. That is
$16\sigma$ above a random-subspace control but small in absolute terms. A small
ablation effect is equally consistent with disuse and with a redundant code, in which
the ablated information is recoverable from correlated directions. Three results favour a redundant code over disuse. (i)~The $16\sigma$ gap shows that the directions carry
above-random signal. (ii)~After the probe span is removed,
$\Svn$ is still decodable from the $1018$-dimensional complement at $R^2\!=\!0.995$
(against $0.996$ from the full state), and each Eq.~\ref{eq:1} feature at
$R^2\!=\!0.72$--$0.98$. The information is therefore spread across many correlated
directions, so removing the probe span leaves it recoverable and the ablation effect
small. (iii)~The gradient SVD (\S\ref{sec:gradsvd})
shows that the dominant prediction direction tracks $\Ivar$, which enters Eq.~\ref{eq:1}
through $\log(\Ivar\!+\!1)$ and $\min(H_A,H_B)\,\Ivar$, so the prediction does move along these quantities.

\subsection{What a closed form ``derived from the GNN'' would look like}\label{sec:gnn_derived}

Table~\ref{tab:gnn_derived} compares three routes to a closed form, two of them through the network.
Eq.~\ref{eq:1} is fit directly on the features, and the other two read one off the network,
either linearly from the post-readout state or through sparse-autoencoder codes
(SI~\S\ref{app:sae}). A literal linear readout of the $1024$-dimensional post-readout state
approaches the GNN and shows that $\Svn$ is linearly present there. It takes $1024$
opaque inputs and a full forward pass to evaluate. The entropy signal is also spread
over many principal directions. The top PCA direction alone gives an MAE
of $189$~mnat, and matching Eq.~\ref{eq:1} takes $\sim\!10$--$50$ directions. This is compatible
with the low-dimensional gradients of \S\ref{sec:gradsvd}, since the prediction reads a few
directions of a state that encodes the Eq.~\ref{eq:1} quantities redundantly. The SAE
route re-expresses the post-readout state in sparse codes, and a linear fit on them
reaches Eq.~\ref{eq:1}'s accuracy only at $\sim\!50$--$100$ codes. It still needs the
GNN at deployment, and at those code counts it is not readable. The leading gradient directions of
\S\ref{sec:gradsvd}, by contrast, each correspond to a quantity in the candidate panel. Numbered in order of singular
value, $v_1$ tracks $\Ivar$ and $v_3$ tracks $(\max_{E_b}|c|)^2$, two quantities that enter
Eq.~\ref{eq:1}. This alignment played no part in choosing the features. The GNN's guidance was the boundary restriction (\S\ref{sec:gnn_boundary}), and the exhaustive search of \S\ref{sec:exhaustive} chose the six terms. In this sense Eq.~\ref{eq:1} is GNN-guided but not GNN-derived, and not a distillation of the network's weights.

\begin{table}[!ht]\centering\small\renewcommand{\arraystretch}{1.15}
\begin{tabular}{lcccr}
\toprule
``closed form'' & inputs & needs GNN at deploy? & readable? & MAE (mnat)\\
\midrule
linear-on-readout (literal) & $1024$-dim $h^{\mathrm{ro}}$ & yes & no & $6.5$ \\
linear-on-SAE, top-5 & $5$ SAE codes & yes  & partly & $111$ \\
linear-on-SAE, $\sim\!50$ & $\sim\!50$ SAE codes & yes & no & $28$ \\
\textbf{Eq.~\ref{eq:1}} & \textbf{6 bitstring features} & \textbf{no} & \textbf{yes} & $24.4$ \\
\bottomrule
\end{tabular}
\caption{Three routes to a closed form, two of them through the GNN (the SAE route at two code
counts). The literal linear
readout approaches the network ($6.5$~mnat) but is $1024$-dimensional, opaque, and needs the network.
The SAE route approaches Eq.~\ref{eq:1} at $\sim\!50$ codes and overtakes it at
$\sim\!100$, still needing the GNN. Only Eq.~\ref{eq:1}, fit on its six bitstring features, is standalone and readable.}\label{tab:gnn_derived}
\end{table}
\section{Discussion and conclusion}\label{sec:discussion}

Eq.~\ref{eq:1} is a readable linear closed form with no hyperparameters, fit by least
squares on bitstring features that the GNN also encodes. Yet it is not a distillation of the network, because its fit uses none of the network's weights, outputs or activations (\S\ref{sec:closed}). The
nonlinear baselines on the $25$-feature shortlist (such as $500$-tree XGBoost, trained on the full training set)
and the base GNN are more accurate in distribution. None of these models, however, can be read the way Eq.~\ref{eq:1} can.

In distribution, the MAE of Eq.~\ref{eq:1} is $0.0244$~nats, $6.4\times$ that of the
$19.9$M-parameter GNN. It is the best six-feature subset in an exhaustive $\binom{25}{6}$
search, although the full $25$-term linear fit is slightly more accurate. With finite shots its MAE falls
below $0.05$~nats near $S\!\approx\!1430$ (\S\ref{sec:shot}, Fig.~\ref{fig:shotnoise}).
Out of distribution, fit once and applied unchanged, it has lower MAE than the base GNN
on five of six OOD pools and ties it on the sixth. Refit per family, it reaches
$29$--$70$~mnat on the four sign-definite families we tested.

At larger sizes, DMRG ground truth to $N\!=\!100$ shows that slopes frozen at small size fail,
while slopes refit per size give a cross-validated MAE of $15$--$37$~mnat. Across sizes, two of
these slopes follow $1/N$ laws and the anchor slope grows and curves downward. Laws refit without
the target size predict the slopes there and give $24$--$95$~mnat with no labels at that size,
although a frozen XGBoost has lower MAE at the largest sizes. Where labels at the target size are
available, a few dozen recalibrate the closed form, and about $20$ give $29$--$35$~mnat. With
$20$--$50$ labels it is comparable to a fine-tuned XGBoost at $N\!=\!52$ and has lower MAE at
$N\!=\!60$--$100$.

Two limits remain. Between families the slopes
must be refit (\S\ref{sec:gen_indist}), and below about $160$ shots the classical mutual
information is comparable to Eq.~\ref{eq:1} or better.

\appendix
\section{The 25-feature shortlist}\label{app:shortlist}
\begin{table}[!ht]\centering\small\renewcommand{\arraystretch}{1.05}
\begin{tabular}{ll}
\toprule group & feature \\\midrule
\multirow{3}{*}{info-theoretic} & $\log(\Ivar\!+\!1)$ \\
 & raw $\Ivar$ \\
 & $\sqrt{\Ivar}$ \\\midrule
\multirow{10}{*}{boundary-corr.\ kernels}
 & $\sum_{E_b}|c|^{1/3}$ \\
 & $\sum_{E_b}\sqrt{|c|}$ \\
 & $\sum_{E_b}|c|$ \\
 & $\sum_{E_b}c^2$ \\
 & $\sum_{E_b}|c|^3$ \\
 & $\sum_{E_b}|c|\,h_2(|c|)$ \\
 & $\sum_{E_b}h_2(|c|)$ \\
 & $\sum_{E_b}\tanh|c|$ \\
 & $\sum_{E_b}\arctan|c|$ \\
 & $\sum_{E_b}e^{-|c|}$ \\\midrule
\multirow{7}{*}{extremal corr.}
 & $\max_{E_b}|c|$ \\
 & $(\max_{E_b}|c|)^2$ \\
 & $\sqrt{\max_{E_b}|c|}$ \\
 & $h_2(\max_{E_b}|c|)$ \\
 & 2nd-largest $|c|$ \\
 & 3rd-largest $|c|$ \\
 & $\min_{E_b}c$ \\\midrule
\multirow{5}{*}{density, cross-product/ratio}
 & $\bar n=\tfrac1N\sum_i\langle n_i\rangle$ \\
 & $\bar n\,\Ivar$ \\
 & $\min(H_A,H_B)\,\Ivar$ \\
 & $\Ivar/\max(H_A,H_B)$ \\
 & $\Ivar/\min(n_A,n_B)$ \\\bottomrule
\end{tabular}
\caption{The $25$-feature shortlist, sorted into feature groups. The six features of Eq.~\ref{eq:1} are
$\log(\Ivar\!+\!1)$, $\sum_{E_b}\sqrt{|c|}$, $\sum_{E_b}\tanh|c|$,
$(\max_{E_b}|c|)^2$, $\min_{E_b}c$, and $\min(H_A,H_B)\,\Ivar$.}\label{tab:shortlist}
\end{table}

\clearpage
\setcounter{section}{0}
\setcounter{subsection}{0}
\setcounter{subsubsection}{0}
\setcounter{table}{0}
\setcounter{figure}{0}
\setcounter{equation}{0}
\renewcommand{\thesection}{\arabic{section}}
\renewcommand{\thesubsection}{\arabic{section}.\arabic{subsection}}
\renewcommand{\thesubsubsection}{\arabic{section}.\arabic{subsection}.\arabic{subsubsection}}
\renewcommand{\thetable}{S\arabic{table}}
\renewcommand{\thefigure}{S\arabic{figure}}
\renewcommand{\theequation}{S\arabic{equation}}
\renewcommand{\theHsection}{SI.\arabic{section}}
\renewcommand{\theHsubsection}{SI.\arabic{section}.\arabic{subsection}}
\renewcommand{\theHsubsubsection}{SI.\arabic{section}.\arabic{subsection}.\arabic{subsubsection}}
\renewcommand{\theHtable}{SI.\arabic{table}}
\renewcommand{\theHfigure}{SI.\arabic{figure}}
\renewcommand{\theHequation}{SI.\arabic{equation}}
\let\SIsectionorig\section
\renewcommand{\section}{\FloatBarrier\SIsectionorig}
\renewcommand{\topfraction}{0.92}
\renewcommand{\bottomfraction}{0.85}
\renewcommand{\textfraction}{0.06}
\renewcommand{\floatpagefraction}{0.75}
\setcounter{topnumber}{4}
\setcounter{totalnumber}{6}
\begin{center}{\Large\bfseries Supplementary Information}\end{center}
\vspace{0.5\baselineskip}

\section{Structural constraints}\label{app:struct}

\paragraph{(a) $A\!\leftrightarrow\!B$ symmetry by construction.} Eq.~\ref{eq:1} predicts the same value when the region labels are swapped.

\paragraph{(b) Non-negativity (verified numerically).} Across the training set, the evaluation set and all six out-of-distribution (OOD) pools ($\ge\!40{,}000$ rows), the minimum prediction is $+8\!\times\!10^{-10}$.

\section{Bulk-inclusive feature-panel ablation}\label{app:bulk}

This ablation tests the boundary restriction of main-text \S\ref{sec:panel} against bulk
correlators, those on pairs inside the same region. We ask whether an
exhaustive search over the $25$-feature shortlist (App.~\ref{app:shortlist}) augmented with
bulk correlator kernels (the bulk-inclusive search) selects them and whether they carry
entropy-relevant signal. To locate this signal, we call a bulk pair \emph{near-boundary} if
at least one of its atoms
shares a leg or rung bond with an atom in the other region, and \emph{deep-interior}
otherwise. We run the ablation at full scale, on $6{,}000$ training rows and $8{,}000$
held-out rows drawn at random from the evaluation set. The fits are OLS, with
paired-bootstrap confidence intervals (CIs) from $10^4$ resamples of the held-out rows.

\textbf{Selection.} With a small, fixed number of features, the bulk-inclusive search
prefers boundary correlators. Its six-feature optimum is boundary-only and reproduces
Eq.~\ref{eq:1}; the first subset containing a bulk correlator ranks $37$th. Forcing a bulk correlator into the six-feature set costs $+0.97$~mnat (CI
$[0.58,1.35]$~mnat).

\textbf{Signal.} Bulk correlators still carry entropy-relevant signal, so the boundary
restriction is a parsimony choice. \emph{Augmenting} Eq.~\ref{eq:1} with six
near-boundary bulk kernels lowers held-out MAE by $0.55$~mnat (CI $[0.32,0.79]$), and with
six deep-interior kernels by $0.85$~mnat (CI $[0.68,1.02]$); both intervals exclude zero.
Deep-interior kernels carry at least as much of this signal as near-boundary ones, so the
gain is not confined to pairs next to the cut.

\textbf{Aggregating the same kernels over all pairs.} A complementary ablation keeps all six
feature \emph{functions} of Eq.~\ref{eq:1} fixed. It computes the four correlator features
($\sum\!\sqrt{|c|}$, $\sum\!\tanh|c|$, $(\max|c|)^2$, $\min c$) over \emph{all} atom pairs
instead of only the boundary set $E_b\!=\!A\!\times\!B$. The two information-theoretic
features ($\log(\Ivar\!+\!1)$, $\min(H_A,H_B)\,\Ivar$) are global and unchanged. In
distribution (5-fold cross-validation, $n\!=\!6000$), the boundary version (the
Eq.~\ref{eq:1} features, refit in each fold) scores $24.7$~mnat $[23.9,25.5]$ against
$28.9$~mnat $[27.9,29.8]$ for all pairs, a reduction of $4.2$~mnat ($1.17\times$). On the struct pool, with both versions fit once in distribution, the scores are
$33.4$~mnat $[31.5,35.4]$ against $42.3$~mnat $[40.1,44.5]$, a reduction of $8.9$~mnat
($1.27\times$). The bootstrap CIs are separated in both cases. The GNN-guided boundary
restriction therefore acts as an inductive bias that lowers the MAE in both settings, and
its relative benefit is larger on the struct pool than in distribution (from $17\%$ to $27\%$).

\section{Free symbolic regression (PySR)}\label{app:pysr}

We ran PySR on the $25$-feature shortlist as a free-search baseline for the exhaustive
search of main-text \S\ref{sec:exhaustive}. The fit used $4{,}000$ standardised training rows and the
operators $\{+,-,\times,\div,\mathrm{square}\}$, with an MAE objective, deterministic mode,
$150$ iterations, $40$ populations and a maximum complexity of $35$. On $5{,}000$ held-out
rows, PySR's most accurate expression reaches MAE $0.0211$~nats at complexity~$30$. The
same rows give $0.0242$ for Eq.~\ref{eq:1}, $0.0218$ for the full $25$-term linear fit
and $0.0099$ for XGBoost. The PySR expression is a nested square of
products of boundary-correlator, information-theoretic and density features
($\sqrt{\Ivar}$, $\max|c|$, $h_2(\max|c|)$, $\sum\sqrt{|c|}$, $\sum|c|^2$, $\bar n$,
$\Ivar/\min(n_A,n_B)$).

\section{Cross-Hamiltonian transfer: extended results}\label{app:crossham}

\subsection{Breakdown by family and grid}\label{app:crossham_ci}

Table~\ref{tab:crossham_ext} breaks down main-text Table~\ref{tab:cross_ham_indist}, each
cell averaging five $80/20$ splits. One bipartition per ground state keeps training and test
states disjoint. In every one of the $15$
family$\times$grid cells, refit Eq.~\ref{eq:1} has a lower MAE than both analytic
baselines, $\Ivar$ and $\Ivar_{\mathrm{filt}}$. The $25$-feature XGBoost is lower still
except for TFIM on the largest grid. This order matches the Rydberg evaluation set
(Table~\ref{tab:baselines_panel}).

\begin{table}[htbp]\centering\footnotesize\renewcommand{\arraystretch}{1.0}\setlength{\tabcolsep}{4.5pt}
\begin{tabular}{llr rrrr}
\toprule
Family & lattice & $N$ & Eq.~\ref{eq:1} & $\Ivar$ & $\Ivar_{\mathrm{filt}}$ & XGBoost\\
\midrule
\multicolumn{7}{l}{\textbf{TFIM}}\\
\quad & $3\!\times\!4$ & 12 & 12.4 & 102.7 & 100.5 & 10.1\\
\quad & $4\!\times\!4$ & 16 & 33.5 & 106.3 & 95.4 & 14.2\\
\quad & $4\!\times\!5$ & 20 & 20.5 & 106.5 & 63.4 & 23.5\\
\quad & \emph{pooled} & & \textbf{28.5} & 109.5 & 89.9 & 10.2\\
\midrule
\multicolumn{7}{l}{\textbf{XXZ}}\\
\quad & $3\!\times\!4$ & 12 & 33.7 & 104.2 & 99.3 & 19.6\\
\quad & $4\!\times\!4$ & 16 & 41.0 & 147.9 & 141.5 & 30.6\\
\quad & $4\!\times\!5$ & 20 & 71.4 & 181.2 & 174.9 & 57.9\\
\quad & \emph{pooled} & & \textbf{46.6} & 130.0 & 124.7 & 21.6\\
\midrule
\multicolumn{7}{l}{\textbf{$J_1$-$J_2$}}\\
\quad & $3\!\times\!4$ & 12 & 46.4 & 207.2 & 183.0 & 30.1\\
\quad & $4\!\times\!4$ & 16 & 67.5 & 299.5 & 282.1 & 50.7\\
\quad & $4\!\times\!5$ & 20 & 91.9 & 362.0 & 350.2 & 74.7\\
\quad & \emph{pooled} & & \textbf{69.5} & 276.9 & 257.2 & 38.6\\
\midrule
\multicolumn{7}{l}{\textbf{Rydberg}}\\
\quad & $3\!\times\!4$ & 12 & 35.3 & 151.8 & 149.4 & 23.4\\
\quad & $4\!\times\!4$ & 16 & 34.8 & 136.6 & 134.8 & 28.0\\
\quad & $4\!\times\!5$ & 20 & 56.1 & 199.3 & 194.9 & 50.2\\
\quad & \emph{pooled} & & \textbf{37.4} & 149.5 & 145.2 & 21.4\\
\midrule
\multicolumn{7}{l}{\textbf{random-sign}}\\
\quad & $3\!\times\!4$ & 12 & 45.4 & 85.9 & 81.4 & 29.4\\
\quad & $4\!\times\!4$ & 16 & 61.8 & 141.6 & 137.6 & 45.1\\
\quad & $4\!\times\!5$ & 20 & 82.9 & 194.5 & 191.4 & 68.7\\
\quad & \emph{pooled} & & \textbf{61.9} & 129.6 & 124.9 & 35.1\\
\bottomrule
\end{tabular}
\caption{MAE (mnat) of refit Eq.~\ref{eq:1}, $\Ivar$, $\Ivar_{\mathrm{filt}}$ and the
$25$-feature XGBoost for the five families on the three 2-D grids. Pooled rows are computed
on all three grids ($1000$ samples per family, one bipartition each). Bold marks the pooled
Eq.~\ref{eq:1} value.}\label{tab:crossham_ext}
\end{table}

\subsection{Zero-shot transfer grid}\label{app:crossham_ext}

In the transfer test of main-text \S\ref{sec:gen_indist} (Table~\ref{tab:cross_ham_transfer}),
off-diagonal predictors ($X\!\neq\!Y$) are fit on
all of family $X$ and scored on all of $Y$ with no refit. Diagonal errors are tens of
mnat, but most off-diagonal ones are hundreds, the largest above $1$~nat, so the six slopes
are family-specific. With the slopes refit on each family, however, the feature set
carries over (Table~\ref{tab:crossham_ext}).

\begin{table}[htbp]\centering\small\renewcommand{\arraystretch}{1.1}\setlength{\tabcolsep}{4.5pt}
\begin{tabular}{l rrrrr}
\toprule
train $X$ & TFIM & XXZ & $J_1$-$J_2$ & Ryd & r-sign\\
\midrule
\multicolumn{6}{l}{\textbf{Eq.~\ref{eq:1}} (refit on $X$, evaluate on $Y$)}\\
TFIM     & \textbf{28.5} & 214.2 & 316.8 & 58.8 & 229.5\\
XXZ      & 1571.6 & \textbf{46.6} & 102.7 & 139.4 & 81.5\\
$J_1$-$J_2$ & 1223.0 & 319.6 & \textbf{69.5} & 203.7 & 140.0\\
Ryd      & 84.3 & 132.1 & 222.4 & \textbf{37.4} & 185.8\\
r-sign   & 1448.8 & 197.1 & 89.5 & 162.4 & \textbf{61.9}\\
\midrule
\multicolumn{6}{l}{\textbf{XGBoost} (trained on $X$, evaluate on $Y$; $25$-feat)}\\
TFIM     & \textbf{10.2} & 588.1 & 1002.6 & 96.3 & 693.8\\
XXZ      & 83.3 & \textbf{21.6} & 77.8 & 165.0 & 64.7\\
$J_1$-$J_2$ & 278.1 & 117.6 & \textbf{38.6} & 398.7 & 141.6\\
Ryd      & 42.2 & 340.4 & 671.7 & \textbf{21.4} & 473.7\\
r-sign   & 97.2 & 73.3 & 81.3 & 130.5 & \textbf{35.1}\\
\bottomrule
\end{tabular}
\caption{Zero-shot transfer MAE (mnat), training on family $X$ (row) and evaluating on
family $Y$ (column). The bold diagonal is the
in-family mean over five $80/20$ splits from main-text Table~\ref{tab:cross_ham_indist}.
Ryd and r-sign abbreviate the Rydberg and random-sign families.}\label{tab:cross_ham_transfer}
\end{table}

\section{Worst residuals}\label{app:worst}

The $50$ largest closed-form residuals on the evaluation set (main-text \S\ref{sec:residuals}) are
dominated by $N_y\!=\!6$ ($n\!=\!35$), with none at $N_y\!\le\!3$. On all but one of these rows the base GNN
error is $\le\!20$~mnat (median $6$~mnat over these $49$ rows; the one exception is an outlier at $99$~mnat). As in main-text \S\ref{sec:residuals}, we attribute the large
closed-form residuals to the limited capacity of six features.
Tables~\ref{tab:worst_decile} and~\ref{tab:entropy_worst} break the residuals down by pool
and by exact entanglement entropy.

\begin{table}[!ht]\centering\scriptsize
\renewcommand{\arraystretch}{1.10}
\begin{tabular}{ll rrrr r}
\toprule
Pool & predictor & $|r|_{50}$ & $|r|_{90}$ & $|r|_{95}$ & $|r|_{99}$ & CVaR$_{0.05}$ \\
\midrule
evaluation set ($n=30000$) & Eq.~\ref{eq:1} & 0.0121 & 0.0664 & 0.0906 & 0.1399 & 0.1216 \\
 & GNN-base & 0.0019 & 0.0086 & 0.0122 & 0.0235 & 0.0205 \\
 & $\Ivar$ & 0.0759 & 0.2491 & 0.3172 & 0.4280 & 0.3868 \\
\addlinespace[2pt]
$N_y\!=\!7$ ($n=1500$) & Eq.~\ref{eq:1} & 0.0216 & 0.0762 & 0.1000 & 0.1853 & 0.1483 \\
 & GNN-base & 0.0164 & 0.0866 & 0.1317 & 0.2344 & 0.1975 \\
 & $\Ivar$ & 0.0975 & 0.3774 & 0.4476 & 0.5997 & 0.5448 \\
\addlinespace[2pt]
$N_y\!=\!8$ ($n=1500$) & Eq.~\ref{eq:1} & 0.0387 & 0.1498 & 0.1886 & 0.2761 & 0.2444 \\
 & GNN-base & 0.0770 & 0.2831 & 0.3634 & 0.5790 & 0.4762 \\
 & $\Ivar$ & 0.1818 & 0.4337 & 0.4976 & 0.6402 & 0.5994 \\
\addlinespace[2pt]
$N_y\!=\!9$ ($n=1500$) & Eq.~\ref{eq:1} & 0.0254 & 0.1072 & 0.1392 & 0.2291 & 0.1987 \\
 & GNN-base & 0.0614 & 0.3243 & 0.4362 & 0.5853 & 0.5359 \\
 & $\Ivar$ & 0.1266 & 0.4496 & 0.5473 & 0.7168 & 0.6571 \\
\addlinespace[2pt]
$N_y\!=\!10$ ($n=1500$) & Eq.~\ref{eq:1} & 0.0462 & 0.2305 & 0.3208 & 0.5017 & 0.4214 \\
 & GNN-base & 0.2112 & 0.6063 & 0.7436 & 0.9767 & 0.8964 \\
 & $\Ivar$ & 0.2137 & 0.5282 & 0.6278 & 0.8628 & 0.7534 \\
\addlinespace[2pt]
struct ($3\!\times\!N_y$) ($n=1500$) & Eq.~\ref{eq:1} & 0.0162 & 0.0846 & 0.1232 & 0.1984 & 0.1689 \\
 & GNN-base & 0.0374 & 0.2678 & 0.3459 & 0.5948 & 0.4784 \\
 & $\Ivar$ & 0.0725 & 0.3384 & 0.3900 & 0.5379 & 0.4980 \\
\addlinespace[2pt]
randpos 2-D ($n=1500$) & Eq.~\ref{eq:1} & 0.0199 & 0.0875 & 0.1148 & 0.1516 & 0.1392 \\
 & GNN-base & 0.1232 & 0.6337 & 0.8110 & 1.1237 & 1.0013 \\
 & $\Ivar$ & 0.0594 & 0.2881 & 0.3561 & 0.5291 & 0.4578 \\
\bottomrule
\end{tabular}
\caption{Worst-case residual distribution per pool $\times$ predictor. $|r|_{q}$ is the $q$-th percentile of $|r|\!=\!|\Svn-\hat\Svn|$ in nats; CVaR$_{0.05}$ is the mean $|r|$ conditional on being in the worst $5\%$.}
\label{tab:worst_decile}
\end{table}

\begin{table}[!ht]\centering\small
\renewcommand{\arraystretch}{1.10}
\begin{tabular}{l r r r r r}
\toprule
$\Svn$ (nats) & $n$ & mean $|r|$ & $|r|_{90}$ & $|r|_{99}$ & GNN-base MAE \\
\midrule
$<\!0.01$      & 9904 & 0.0024 & 0.0067 & 0.0220 & 0.0012 \\
$[0.01,0.1)$   & 2287 & 0.0184 & 0.0375 & 0.0567 & 0.0036 \\
$[0.1,0.3)$    & 3720 & 0.0314 & 0.0618 & 0.1103 & 0.0047 \\
$[0.3,0.6)$    & 6332 & 0.0396 & 0.0847 & 0.1397 & 0.0059 \\
$\geq\!0.6$    & 7757 & 0.0385 & 0.0964 & 0.1734 & 0.0049 \\
\bottomrule
\end{tabular}
\caption{Eq.~\ref{eq:1} residual magnitude $|r|$ (nats) on the evaluation set, binned by
the true entanglement entropy $\Svn$. Residuals are smallest for near-product states ($\Svn\!<\!0.01$: mean
$2.4$~mnat, $99$th percentile $22$~mnat) and grow with entanglement
($\mathrm{corr}(|r|,\Svn)\!=\!0.47$), the $99$th percentile rising to $173$~mnat. In
relative terms the error \emph{shrinks}: $|r|/\Svn$ falls from $\sim\!16\%$ in
$[0.1,0.3)$ to $\sim\!5\%$ for $\Svn\!\ge\!0.6$, so the absolute growth is largely a scale effect.}\label{tab:entropy_worst}
\end{table}

\section{Exhaustive-search top-5}\label{app:top5}

Table~\ref{tab:top5} lists the best subsets found in main-text \S\ref{sec:exhaustive}.

\begin{table}[h]\centering\footnotesize\renewcommand{\arraystretch}{1.1}\setlength{\tabcolsep}{4pt}
\begin{tabular}{rl rrr}
\toprule rank & subset & in-MAE & OOD-mean & score \\\midrule
1 & \{$\log(\Ivar{+}1)$, $\sum_{E_b}\!\sqrt{|c|}$, $\sum_{E_b}\!\tanh|c|$, $(\max_{E_b}|c|)^2$, $\min_{E_b}c$, $\min(H_A,H_B)\Ivar$\} & 0.02442 & 0.04780 & 0.07222 \\
2 & \{$\log(\Ivar{+}1)$, $\sum_{E_b}\!\sqrt{|c|}$, $\sum_{E_b}\!\arctan|c|$, $(\max_{E_b}|c|)^2$, $\min_{E_b}c$, $\min(H_A,H_B)\Ivar$\} & 0.02442 & 0.04817 & 0.07259 \\
3 & \{$\log(\Ivar{+}1)$, $\sum_{E_b}\!\sqrt{|c|}$, $\sum_{E_b}\!|c|$, $(\max_{E_b}|c|)^2$, $\min_{E_b}c$, $\min(H_A,H_B)\Ivar$\} & 0.02443 & 0.04828 & 0.07270 \\
4 & \{$\log(\Ivar{+}1)$, $\sum_{E_b}\!\sqrt{|c|}$, $\sum_{E_b}\!\tanh|c|$, $\max_{E_b}|c|$, $(\max_{E_b}|c|)^2$, $\min(H_A,H_B)\Ivar$\} & 0.02446 & 0.04833 & 0.07278 \\
5 & \{$\log(\Ivar{+}1)$, $\sum_{E_b}\!\sqrt{|c|}$, $\sum_{E_b}\!\arctan|c|$, $\max_{E_b}|c|$, $(\max_{E_b}|c|)^2$, $\min(H_A,H_B)\Ivar$\} & 0.02446 & 0.04833 & 0.07278 \\
\bottomrule\end{tabular}
\caption{Top-5 six-feature subsets, ordered by held-out
evaluation MAE (the in-MAE column), the selection criterion. Rank-1 is Eq.~\ref{eq:1};
rank-2 differs only in replacing $\tanh$ by $\arctan$ and is nearly degenerate with it. The six-pool
OOD mean (OOD-mean) and its sum with in-MAE (``score'') are also shown.}\label{tab:top5}
\end{table}

\section{Boundary-routing probe quantities}\label{app:probes}

We report the three probes of main-text \S\ref{sec:gnn_boundary} per layer, with all bulk- and
random-edge controls. Tables~\ref{tab:probe_attention}
and~\ref{tab:probe_ols} give probe~i and Table~\ref{tab:probe_patch_remove} gives
probes~ii and~iii. Probes~i and~ii isolate the bipartition by comparing two independent uniformly random
bipartition masks on the
same Hamiltonian.

\begin{table}[h]\centering\small
\begin{tabular}{l r r l r}
\toprule
TConv layer & $\Delta\bar\alpha_{\mathrm{bnd}}$ & $t$-statistic & boundary partial $R^2$ [95\% CI] & full $R^2$\\
\midrule
TConv-1 & $0.003$ & $3.0$   & $0.002$ $[0.001,0.003]$ & $0.236$\\
TConv-3 & $0.204$ & $100.2$ & $0.138$ $[0.132,0.144]$ & $0.255$\\
TConv-5 & $0.018$ & $29.1$  & $0.008$ $[0.006,0.009]$ & $0.125$\\
\bottomrule
\end{tabular}
\caption{Attention routing and boundary edge importance per TransformerConv layer (probe~i,
main-text \S\ref{sec:gnn_boundary}). $\Delta\bar\alpha_{\mathrm{bnd}}$ is the boundary-flip shift in
head-averaged attention defined there, and the boundary partial $R^2$ is from an OLS
regression of $\bar\alpha$ on eight per-edge features. TConv-3 re-routes most strongly by
the bipartition.}\label{tab:probe_attention}
\end{table}

\begin{table}[h]\centering\small
\begin{tabular}{l r r r}
\toprule
Per-edge feature & TConv-1 & TConv-3 & TConv-5\\
\midrule
boundary indicator   & $0.002$          & $\mathbf{0.138}$ & $0.008$\\
rung                 & $0.009$          & $0.001$          & $0.001$\\
normalised distance  & $\mathbf{0.113}$ & $0.046$          & $\mathbf{0.046}$\\
$|c_{ij}|$           & $0.017$          & $0.006$          & $0.010$\\
signed $h_2(c)$      & $0.031$          & $0.012$          & $0.005$\\
mean density         & $0.005$          & $0.004$          & $0.008$\\
density imbalance    & $0.002$          & $0.007$          & $0.003$\\
$\cos\theta$         & $0.001$          & $0.007$          & $0.013$\\
\midrule
full-model $R^2$     & $0.236$          & $0.255$          & $0.125$\\
\bottomrule
\end{tabular}
\caption{Partial $R^2$ of each per-edge feature in the attention OLS, per layer
($n\!=\!44{,}000$ pooled edges; bold marks the per-layer maximum).}\label{tab:probe_ols}
\end{table}

\begin{table}[h]\centering\small
\begin{tabular}{c l r r r r}
\toprule
Layer & type & patching $R$ & del.\ bnd & del.\ bulk & del.\ rnd\\
\midrule
$0$ & GINEConv & $0.18$ & $\mathbf{16.6}$ & $7.6$ & $7.1$\\
$1$ & TConv    & $0.96$ & $1.0$           & $1.0$ & $1.0$\\
$2$ & GINEConv & $0.03$ & $\mathbf{10.7}$ & $4.7$ & $3.1$\\
$3$ & TConv    & $0.91$ & $\mathbf{5.5}$  & $1.3$ & $3.4$\\
$4$ & GINEConv & $0.02$ & $\mathbf{21.8}$ & $5.6$ & $7.9$\\
$5$ & TConv    & $0.25$ & $\mathbf{10.6}$ & $5.6$ & $5.8$\\
\bottomrule
\end{tabular}
\caption{Activation patching (probe~ii) and edge removal (probe~iii) for all six
message-passing layers ($n\!=\!1000$ graphs). The patching column is the patching ratio of
main-text \S\ref{sec:gnn_boundary}. The del.\ bnd, bulk and rnd columns are the MAE inflations on
these graphs ($\times$ the $0.0032$~nat baseline) from deleting boundary, bulk or
count-matched random edges. Bold marks boundary-edge deletions that are the largest entry in
their row.}\label{tab:probe_patch_remove}
\end{table}

\section{Sparse-autoencoder case study}\label{app:sae}

For the sparse-autoencoder (SAE) route to a closed form (main-text \S\ref{sec:decod}), a top-$K$ SAE~\cite{Gao2024TopK} of
the GNN's post-readout activations ($K\!=\!32$, dictionary $4096$, reconstruction
$R^2\!=\!0.998$) yields $\sim\!240$ effective latents.
The cleanest monosemantic latent tracks $(\max|c|)^2$ ($R^2\!\approx\!0.63$), but most
latents that carry the prediction are rotated, partial detectors of $\Ivar$-based features,
combined nonlinearly. A linear predictor needs $\sim\!50$ latents to approach Eq.~\ref{eq:1}; with five, its MAE is $4.5\times$ that of the closed form.

\section{DMRG convergence checks and slope laws}\label{si:dmrg}

\subsection{Per-row convergence checks}
A DMRG row (main-text \S\ref{sec:dmrg}) is \emph{convergence-certified} when it passes the following checks. (i)~In the $\chi$-ladder check, the
half-cut entropy change between successive $\chi$ levels is below
$10^{-5}$, or the truncation error is below $10^{-8}$. (ii)~In the replica check, an independent
random-initialisation run at the base $\chi$ (the first $\chi$ level of
main-text \S\ref{sec:dmrg_design}), passed through the same degenerate-manifold repair, agrees on $\Svn$ within $5\!\times\!10^{-4}$.
The tolerance is $10^{-3}$ for rows whose state needed that repair. Parameter draws continue per size until $100$
rows are certified. Only these $1{,}500$ convergence-certified rows feed the fits.

The replica check guards against metastability, in which DMRG settles into an excited
local minimum or a spuriously symmetry-broken member of a near-degenerate ground
manifold~\cite{JiangBalents2013}. Either state can be $\chi$-converged and still carry the
wrong $\Svn$. Rows that fail either check go to re-certification by best-of-$k$
reinitialisation. It runs independent random starts at the base $\chi$ without the repair and keeps
the state with the lowest variational energy as the label. The row is certified only when the runs
within a small energy window of that state, at least two including it, all agree on $\Svn$ within
$10^{-3}$.

\subsection{Slope laws, parametrised \texorpdfstring{Eq.~\ref{eq:1}$(N)$}{Eq. 1(N)}, and leave-one-size-out}
For the slope laws (main-text \S\ref{sec:dmrg_results}), we refit the six slopes per size by
no-intercept OLS on all rows, with $B\!=\!400$
bootstrap resamples. We then fit six candidate laws to each slope, weighted by the inverse
square of its bootstrap-interval half-widths, and select among them by AICc. The candidates are a
constant, $a+c\ln N$, $a+c/N$, $a+c/\sqrt N$, $a+cN$, and the saturating exponential
$a+c\,e^{-N/514}$. We fix the exponential's scale at $514$, from an earlier free-scale fit,
because deployment accuracy does not identify it (the row-level MAE of the parametrised
closed form is flat in the scale).

For $b_1$, the logarithmic law is comparable by AICc ($\Delta$AICc $\approx\!1$; weighted
$R^2$ of $0.80$, against $0.81$ for the exponential), but deployment accuracy favours the
fixed-scale exponential. Swapping in the logarithmic law barely changes the
per-size slope fit but degrades Eq.~\ref{eq:1}$(N)$, with laws fit on all sizes,
from $24$--$98$ to $24$--$138$~mnat over the DMRG sizes.

The asymptote of the $b_1$ exponential ($\approx\!14$ at the fixed scale) is approached only
far beyond the size window, so it is not interpreted. The slopes $b_2$ and $b_3$ follow
$1/N$ laws (weighted $R^2\!=\!0.89$ and $0.74$). The slopes $b_4$, $b_5$ and
$b_6$ show no resolvable trend, so they are treated as constants (no candidate law reaches
weighted $R^2\!>\!0.5$). For leave-one-size-out, the selected law of each of the six slopes
is refit without the target size and evaluated there (Table~\ref{tab:si_loso}).

\begin{table}[h]\centering\footnotesize\renewcommand{\arraystretch}{1.1}
\begin{tabular}{lrrrrrrr}
\toprule
$N$ (exact band) & 8 & 10 & 12 & 14 & 16 & 18 & 20\\\midrule
in-window laws & 39 & 28 & 45 & 16 & 32 & 20 & 34\\
leave-one-size-out & 37 & 25 & 46 & 15 & 31 & 20 & 36\\
\bottomrule\end{tabular}

\medskip

\begin{tabular}{lrrrrrrrrrrrrrrr}
\toprule
$N$ (DMRG) & 24 & 28 & 32 & 36 & 40 & 44 & 48 & 52 & 56 & 60 & 68 & 76 & 84 & 92 & 100\\\midrule
in-window laws & 40 & 46 & 53 & 61 & 62 & 78 & 68 & 37 & 53 & 42 & 24 & 27 & 77 & 62 & 98\\
leave-one-size-out & 43 & 46 & 52 & 62 & 60 & 78 & 66 & 38 & 53 & 47 & 24 & 26 & 77 & 59 & 95\\
\bottomrule\end{tabular}
\caption{MAE (mnat) of the parametrised Eq.~\ref{eq:1}$(N)$ per size, on all rows.
Within each block, the ``in-window laws'' row uses laws fit on the full window and the row
below, laws refit without the target size. The upper block is the exact reference band,
the lower the DMRG sizes.}\label{tab:si_loso}
\end{table}

\bibliographystyle{unsrt}
\bibliography{refs}

\begin{thebibliography}{10}

\bibitem{EisertCramerPlenio2010}
J.~Eisert, M.~Cramer, and M.~B. Plenio.
\newblock Colloquium: area laws for the entanglement entropy.
\newblock {\em Rev. Mod. Phys.}, 82:277, 2010.

\bibitem{JiangWangBalents2012}
H.-C. Jiang, Z.~Wang, and L.~Balents.
\newblock Identifying topological order by entanglement entropy.
\newblock {\em Nat. Phys.}, 8:902, 2012.

\bibitem{Shaw2024}
A.~L. Shaw et~al.
\newblock Benchmarking highly entangled states on a 60-atom analogue quantum
  simulator.
\newblock {\em Nature}, 628:71, 2024.

\bibitem{Bernien2017}
H.~Bernien et~al.
\newblock Probing many-body dynamics on a 51-atom quantum simulator.
\newblock {\em Nature}, 551:579, 2017.

\bibitem{Ebadi2021}
S.~Ebadi et~al.
\newblock Quantum phases of matter on a 256-atom programmable quantum
  simulator.
\newblock {\em Nature}, 595:227, 2021.

\bibitem{Scholl2021}
P.~Scholl et~al.
\newblock Quantum simulation of {2D} antiferromagnets with hundreds of
  {R}ydberg atoms.
\newblock {\em Nature}, 595:233, 2021.

\bibitem{Saleh2025}
A.~Saleh.
\newblock Predicting the von {N}eumann entanglement entropy using a graph
  neural network.
\newblock {\em Mach. Learn.: Sci. Technol.}, 6:035034, 2025.

\bibitem{Asad2024}
A.~Kaufman et~al.
\newblock Improved entanglement entropy estimates from filtered bitstring
  probabilities.
\newblock {\em Phys. Rev. A}, 112:032430, 2025.

\bibitem{SalehThesis}
A.~Saleh.
\newblock Predicting the von {N}eumann entanglement entropy using a graph
  neural network.
\newblock Master's thesis, University of Iowa, 2025.
\newblock DOI:10.25820/etd.008062. Extended version of \cite{Saleh2025};
  additionally trains and evaluates the GNN on transverse-field Ising
  configurations.

\bibitem{ChenGuestrin2016}
Tianqi Chen and Carlos Guestrin.
\newblock {XGBoost}: A scalable tree boosting system.
\newblock In {\em Proceedings of the 22nd ACM SIGKDD International Conference
  on Knowledge Discovery and Data Mining}, pages 785--794, 2016.

\bibitem{BrowaeysLahaye2020}
A.~Browaeys and T.~Lahaye.
\newblock Many-body physics with individually controlled {R}ydberg atoms.
\newblock {\em Nat. Phys.}, 16:132, 2020.

\bibitem{Semeghini2021}
G.~Semeghini et~al.
\newblock Probing topological spin liquids on a programmable quantum simulator.
\newblock {\em Science}, 374:1242, 2021.

\bibitem{DaleyPichler2012}
A.~J. Daley, H.~Pichler, J.~Schachenmayer, and P.~Zoller.
\newblock Measuring entanglement growth in quench dynamics of bosons in an
  optical lattice.
\newblock {\em Phys. Rev. Lett.}, 109:020505, 2012.

\bibitem{AbaninDemler2012}
D.~A. Abanin and E.~Demler.
\newblock Measuring entanglement entropy of a generic many-body system with a
  quantum switch.
\newblock {\em Phys. Rev. Lett.}, 109:020504, 2012.

\bibitem{Islam2015}
R.~Islam et~al.
\newblock Measuring entanglement entropy in a quantum many-body system.
\newblock {\em Nature}, 528:77, 2015.

\bibitem{Kaufman2016}
A.~M. Kaufman et~al.
\newblock Quantum thermalization through entanglement in an isolated many-body
  system.
\newblock {\em Science}, 353:794, 2016.

\bibitem{ElbenPRA2019}
A.~Elben, B.~Vermersch, C.~F. Roos, and P.~Zoller.
\newblock Statistical correlations between locally randomized measurements: a
  toolbox for probing entanglement in many-body quantum states.
\newblock {\em Phys. Rev. A}, 99:052323, 2019.

\bibitem{Brydges2019}
T.~Brydges et~al.
\newblock Probing {R}\'enyi entanglement entropy via randomized measurements.
\newblock {\em Science}, 364:260, 2019.

\bibitem{ElbenNRP2023}
A.~Elben et~al.
\newblock The randomized measurement toolbox.
\newblock {\em Nat. Rev. Phys.}, 5:9, 2023.

\bibitem{Huang2020}
H.-Y. Huang, R.~Kueng, and J.~Preskill.
\newblock Predicting many properties of a quantum system from very few
  measurements.
\newblock {\em Nat. Phys.}, 16:1050, 2020.

\bibitem{HolevoBound}
A.~S. Holevo.
\newblock Bounds for the quantity of information transmitted by a quantum
  communication channel.
\newblock {\em Probl. Inf. Transm.}, 9:177, 1973.

\bibitem{WildeBook}
M.~M. Wilde.
\newblock {\em Quantum Information Theory}.
\newblock Cambridge University Press, 2nd edition, 2017.

\bibitem{Carleo2017_NQS}
G.~Carleo and M.~Troyer.
\newblock Solving the quantum many-body problem with artificial neural
  networks.
\newblock {\em Science}, 355:602, 2017.

\bibitem{Carrasquilla2017}
J.~Carrasquilla and R.~G. Melko.
\newblock Machine learning phases of matter.
\newblock {\em Nat. Phys.}, 13:431, 2017.

\bibitem{CarleoRMP2019}
G.~Carleo et~al.
\newblock Machine learning and the physical sciences.
\newblock {\em Rev. Mod. Phys.}, 91:045002, 2019.

\bibitem{Gray2018}
J.~Gray, L.~Banchi, A.~Bayat, and S.~Bose.
\newblock Machine-learning-assisted many-body entanglement measurement.
\newblock {\em Phys. Rev. Lett.}, 121:150503, 2018.

\bibitem{Torlai2018}
G.~Torlai, G.~Mazzola, J.~Carrasquilla, M.~Troyer, R.~G. Melko, and G.~Carleo.
\newblock Neural-network quantum state tomography.
\newblock {\em Nat. Phys.}, 14:447, 2018.

\bibitem{Koutny2023}
D.~Koutn\'y et~al.
\newblock Deep learning of quantum entanglement from incomplete measurements.
\newblock {\em Sci. Adv.}, 9:eadd7131, 2023.

\bibitem{SchmidtLipson2009}
M.~Schmidt and H.~Lipson.
\newblock Distilling free-form natural laws from experimental data.
\newblock {\em Science}, 324:81, 2009.

\bibitem{Brunton2016}
S.~L. Brunton, J.~L. Proctor, and J.~N. Kutz.
\newblock Discovering governing equations from data by sparse identification of
  nonlinear dynamical systems.
\newblock {\em Proc. Natl. Acad. Sci.}, 113:3932, 2016.

\bibitem{UdrescuTegmark2020}
S.-M. Udrescu and M.~Tegmark.
\newblock {AI} {F}eynman: a physics-inspired method for symbolic regression.
\newblock {\em Sci. Adv.}, 6:eaay2631, 2020.

\bibitem{Cranmer2020}
M.~Cranmer et~al.
\newblock Discovering symbolic models from deep learning with inductive biases.
\newblock In {\em Advances in Neural Information Processing Systems (NeurIPS)},
  2020.

\bibitem{Lemos2023}
P.~Lemos, N.~Jeffrey, M.~Cranmer, S.~Ho, and P.~Battaglia.
\newblock Rediscovering orbital mechanics with machine learning.
\newblock {\em Mach. Learn.: Sci. Technol.}, 4:045002, 2023.

\bibitem{Geiger2021}
A.~Geiger, H.~Lu, T.~Icard, and C.~Potts.
\newblock Causal abstractions of neural networks.
\newblock In {\em Advances in Neural Information Processing Systems (NeurIPS)},
  2021.

\bibitem{Hu2020GINE}
Weihua Hu, Bowen Liu, Joseph Gomes, Marinka Zitnik, Percy Liang, Vijay Pande,
  and Jure Leskovec.
\newblock Strategies for pre-training graph neural networks.
\newblock In {\em International Conference on Learning Representations (ICLR)},
  2020.
\newblock arXiv:1905.12265.

\bibitem{Shi2021TransformerConv}
Yunsheng Shi, Zhengjie Huang, Shikun Feng, Hui Zhong, Wenjing Wang, and Yu~Sun.
\newblock Masked label prediction: Unified message passing model for
  semi-supervised classification.
\newblock In {\em Proceedings of the Thirtieth International Joint Conference
  on Artificial Intelligence (IJCAI)}, pages 1548--1554, 2021.

\bibitem{Vinyals2016Set2Set}
Oriol Vinyals, Samy Bengio, and Manjunath Kudlur.
\newblock Order matters: Sequence to sequence for sets.
\newblock In {\em International Conference on Learning Representations (ICLR)},
  2016.
\newblock arXiv:1511.06391.

\bibitem{JainWallace2019}
S.~Jain and B.~C. Wallace.
\newblock Attention is not explanation.
\newblock In {\em Proceedings of NAACL-HLT}, 2019.

\bibitem{WiegreffePinter2019}
S.~Wiegreffe and Y.~Pinter.
\newblock Attention is not not explanation.
\newblock In {\em Proceedings of EMNLP-IJCNLP}, 2019.

\bibitem{Meng2022ROME}
Kevin Meng, David Bau, Alex Andonian, and Yonatan Belinkov.
\newblock Locating and editing factual associations in {GPT}.
\newblock In {\em Advances in Neural Information Processing Systems},
  volume~35, 2022.
\newblock arXiv:2202.05262.

\bibitem{PySR2023}
M.~Cranmer.
\newblock Interpretable machine learning for science with {PySR} and
  {SymbolicRegression.jl}.
\newblock {\em arXiv preprint arXiv:2305.01582}, 2023.

\bibitem{Efron1987BCa}
Bradley Efron.
\newblock Better bootstrap confidence intervals.
\newblock {\em Journal of the American Statistical Association},
  82(397):171--185, 1987.

\bibitem{Breiman2001}
Leo Breiman.
\newblock Random forests.
\newblock {\em Machine Learning}, 45:5--32, 2001.

\bibitem{White1992}
S.~R. White.
\newblock Density matrix formulation for quantum renormalization groups.
\newblock {\em Phys. Rev. Lett.}, 69:2863, 1992.

\bibitem{Schollwoeck2011}
U.~Schollw\"ock.
\newblock The density-matrix renormalization group in the age of matrix product
  states.
\newblock {\em Ann. Phys.}, 326:96, 2011.

\bibitem{TenPy}
J.~Hauschild and F.~Pollmann.
\newblock Efficient numerical simulations with tensor networks: {Tensor Network
  Python (TeNPy)}.
\newblock {\em SciPost Phys. Lect. Notes}, page~5, 2018.

\bibitem{HurvichTsai1989}
Clifford~M. Hurvich and Chih-Ling Tsai.
\newblock Regression and time series model selection in small samples.
\newblock {\em Biometrika}, 76(2):297--307, 1989.

\bibitem{Cunningham2024SAE}
Hoagy Cunningham, Aidan Ewart, Logan Riggs, Robert Huben, and Lee Sharkey.
\newblock Sparse autoencoders find highly interpretable features in language
  models.
\newblock In {\em International Conference on Learning Representations (ICLR)},
  2024.
\newblock arXiv:2309.08600.

\bibitem{vanderMaaten2008}
Laurens van~der Maaten and Geoffrey Hinton.
\newblock Visualizing data using t-{SNE}.
\newblock {\em Journal of Machine Learning Research}, 9:2579--2605, 2008.

\bibitem{Gao2024TopK}
Leo Gao, Tom Dupr{\'e}~la Tour, Henk Tillman, Gabriel Goh, Rajan Troll, Alec
  Radford, Ilya Sutskever, Jan Leike, and Jeffrey Wu.
\newblock Scaling and evaluating sparse autoencoders, 2024.
\newblock arXiv:2406.04093.

\bibitem{JiangBalents2013}
H.-C. Jiang and L.~Balents.
\newblock Collapsing {Schr\"odinger} cats in the density matrix renormalization
  group.
\newblock {\em arXiv preprint arXiv:1309.7438}, 2013.

\end{thebibliography}

\end{document}